\documentclass{aa}

\usepackage[pdfpagelabels=false]{hyperref}
\hypersetup{colorlinks=true,
            linkcolor=blue,
            citecolor=blue,
            filecolor=blue,
            urlcolor=blue,
            }

\usepackage{graphicx}	
\usepackage{amsmath}	
\usepackage{amssymb}	
\usepackage{multicol}        
\usepackage{bm}		
\usepackage{pdflscape}	
\usepackage{xcolor}

\newcommand{\hGpc}{{\ifmmode{h^{-1}{\rm Gpc}}\else{$h^{-1}$Gpc}\fi}}
\newcommand{\hMpc}{{\ifmmode{h^{-1}{\rm Mpc}}\else{$h^{-1}$Mpc}\fi}}
\newcommand{\hkpc}{{\ifmmode{h^{-1}{\rm kpc}}\else{$h^{-1}$kpc}\fi}}
\newcommand{\hMsun}{{\ifmmode{h^{-1}{\rm {M_{\odot}}}}\else{$h^{-1}{\rm{M_{\odot}}}$}\fi}}
\newcommand{\Mstar}{{\ifmmode{M_{*}}\else{$M_{*}$}\fi}}
\newcommand{\Mhalo}{{\ifmmode{M_{\rm Halo}}\else{$M_{\rm Halo}$}\fi}}
\newcommand{\Ngal}{{\ifmmode{N_{\rm gal}}\else{$N_{\rm gal}$}\fi}}
\newcommand{\Norph}{{\ifmmode{N_{\rm orphan}}\else{$N_{\rm orphan}$}\fi}}
\newcommand{\Nxorph}{{\ifmmode{N_{\rm non-orphan}}\else{$N_{\rm non-orphan}$}\fi}}
\newcommand{\Zsolar}{{\ifmmode{Z_{\odot}}\else{$Z_{\odot}$}\fi}}
\newcommand{\Msun}{{\ifmmode{{\rm {M_{\odot}}}}\else{${\rm{M_{\odot}}}$}\fi}}
\newcommand{\ltsima}{$\; \buildrel < \over \sim \;$}
\newcommand{\gtsima}{$\; \buildrel > \over \sim \;$}
\newcommand{\lsim}{\lower.5ex\hbox{\ltsima}}
\newcommand{\gsim}{\lower.5ex\hbox{\gtsima}}

\newcommand{\Tab}[1]{Table~\ref{#1}}
\newcommand{\Sec}[1]{Section~\ref{#1}}
\newcommand{\App}[1]{Appendix~\ref{#1}}

\newcommand{\Fig}[1]{Fig.~\ref{#1}}
\newcommand{\beq}{\begin{equation}}
\newcommand{\eeq}{\end{equation}}

\usepackage[T1]{fontenc}
\usepackage{ae,aecompl}

\usepackage{newtxtext,newtxmath}
\usepackage{natbib}
\bibpunct{(}{)}{;}{a}{}{,} 

\begin{document}

   \titlerunning{Characterising the ICL in The300}
   \title{Characterising the intra-cluster light in The Three Hundred simulations}

   \author{A. Contreras-Santos
          \inst{1}
          \and
          A. Knebe \inst{1,2,3}
          \and
          W. Cui \inst{1,2,4}
          \and
          I. Alonso Asensio \inst{5,6}
          \and
          C. Dalla Vecchia \inst{5,6}
          \and
          R. Cañas \inst{1}
          \and
          R. Haggar \inst{7,8}
          \and
          R. A. Mostoghiu Paun \inst{9,10}
          \and
          F. R. Pearce \inst{11}
          \and
          E. Rasia \inst{12,13}
          }

   \institute{Departamento de F\'isica Te\'{o}rica, M\'{o}dulo 15, Facultad de Ciencias, Universidad Aut\'{o}noma de Madrid, 28049 Madrid, Spain \\
              \email{ana.contreras@uam.es}
         \and
             Centro de Investigaci\'{o}n Avanzada en F\'isica Fundamental (CIAFF), Facultad de Ciencias, Universidad Aut\'{o}noma de Madrid, 28049 Madrid, Spain
         \and
             International Centre for Radio Astronomy Research, University of Western Australia, 35 Stirling Highway, Crawley, Western Australia 6009, Australia
         \and
             Institute for Astronomy, University of Edinburgh, Royal Observatory, Blackford Hill, Edinburgh EH9 3HJ, UK
         \and
             Instituto de Astrof\'{i}sica de Canarias, C/V\'{i}a L\'{a}ctea s/n, 38205 La Laguna, Tenerife, Spain
         \and
             Departamento de Astrof\'{i}sica, Universidad de La Laguna, Av. Astrof\'{i}sico Francisco Sanchez s/n, E-38206 La Laguna, Tenerife, Spain
         \and
             Department of Physics and Astronomy, University of Waterloo, Waterloo, Ontario N2L 3G1, Canada
        \and
             Waterloo Centre for Astrophysics, University of Waterloo, Waterloo, Ontario N2L 3G1, Canada
        \and
            Centre for Astrophysics \& Supercomputing, Swinburne University of Technology, 1 Alfred St, Hawthorn, VIC 3122, Australia
        \and
            ARC Centre of Excellence for Dark Matter Particle Physics (CDM), Australia
        \and
            School of Physics \& Astronomy, University of Nottingham, Nottingham NG7 2RD, United Kingdom
        \and
            INAF – Osservatorio Astronomico di Trieste, via Tiepolo 11, I34131 Trieste, Italy
        \and
            IFPU – Institute for Fundamental Physics of the Universe, via Beirut 2, 34151, Trieste, Italy
             }

\date{Received January 1, 2020; accepted January 1, 2021}

 
  \abstract{We characterise the intra-cluster light (ICL) in ensembles of full-physics cluster simulations from \textsc{The Three Hundred} project, a suite of 324 hydrodynamical resimulations of cluster-sized halos. We identify the ICL as those stellar particles bound to the potential of the cluster itself, but not to any of its substructures, and separate the brightest cluster galaxy (BCG) by means of a fixed 50 kpc aperture. We find the total BCG+ICL mass to be in agreement with state-of-the-art observations of galaxy clusters. The ICL mass fraction of our clusters is between 30 and 50 per cent of the total stellar mass within $R_{500}$, while the BCG represents around 10 percent. We further find no trend of the ICL fraction with cluster halo mass, at least not in the range $[0.2,3]10^{15}\hMsun$ considered here. For the dynamical state, characterised both by theoretical estimators and by the recent merging history of the cluster, there is a clear correlation, such that more relaxed clusters and those that have undergone fewer recent mergers have a higher ICL fraction. Finally, we investigate the possibility of using the ICL to explore the dark matter (DM) component of galaxy clusters. We compute the volumetric density profile for the DM and ICL components and show that, up to $R_{500}$, the ratio between the two can be described by a power law. Working with the velocity dispersion profiles instead, we show that the ratio can be fit by a straight line. Providing the parameters of these fits, we show how the ICL can be used to infer DM properties.}


   \maketitle
%

\section{Introduction}

Galaxy clusters are the most massive objects in the Universe. Their properties are the result of the complex interplay of multi-scale physical processes and the hierarchical assembly of numerous galaxies. A key component of these objects is the so-called intra-cluster light (ICL) coming from a diffuse distribution of stars. This stellar component was first theorised by \citet{Zwicky1937} and was later confirmed by observations of the nearby Coma cluster \citep{Zwicky1951}. It is likely the by-product of the numerous interactions between galaxies during the assembly and continuous growth of clusters. The study of this diffuse stellar component can therefore give us important information about the assembly history and dynamical state of galaxy clusters and the most massive galaxies in the Universe \citep[e.g.][]{Contini2021,Montes2022,Ragusa2023,Contini2023a}.

In recent years, there has been growing interest from the scientific community to study the ICL both from observations \citep[e.g.][]{Furnell2021,deOliveira2022,Werner2023,Zhang2023} and theory \citep[e.g.][]{Chun2022,Chun2023,Marini2022,Tang2023}. In observations, this is motivated by the increasing sensitivity of instruments and telescopes, as well as improvements in image-processing techniques. The availability of very deep observations, both from ground-based facilities and from space telescopes has allowed groundbreaking observations such as the detection of ICL at $z>1$ using deep infrared imaging data from the Hubble Space Telescope \citep{Joo2023}, and the first ICL study of a cluster with JWST data \citep{MontesTrujillo2022}. On the side of simulations, there is also increasing interest due to the availability of more computational resources and the progress in understanding and modelling of the subgrid processes responsible for the formation of stars and their feedback. Cosmological simulations like IllustrisTNG evolve a large enough volume to study the evolution of massive clusters and galaxies \citep{Pillepich2018a,Pillepich2018b}, while at the same time resolving the ICL component, too.

In order to study the ICL, the first step is to have some way to define it, which is non-trivial and therefore one of the most important problems in the field\footnote{The names used for ICL from different studies can also differ, it can be called diffuse stellar component (DSC), or intra-halo stellar component (IHSC). In this paper, for simplicity we will refer to it as ICL.}. Theoretically, the ICL can be defined as stars that are not bound to any particular galaxy but to the potential of the cluster itself. However, in practice, applying this definition in observations is extremely difficult. Common approaches include surface brightness cuts (e.g. \citealp{Rudick2011,Feldeimer2004,Burke2015}) and fitting of functional forms such as Sérsic (e.g. \citealp{Gonzalez2005,Janowiecki2010,Spavone2018}) or Navarro-Frenk-White profiles \citep{ContiniGu2020,ContiniGu2021}. For a more in-depth discussion of this matter, exploring the advantages and disadvantages of each definition, the reader is referred to the reviews by \citet{Montes2019,Montes2022} and \citet{Contini2021}.

In simulations, where phase-space information is available, the task of identifying the ICL still poses some issues in separating the different gravitational potentials of galaxies and host clusters \citep{Canas2020}. Some studies adopt the approach of considering the joint system of the brightest cluster galaxy (BCG) and the ICL surrounding it, rather than trying to separate one component from the other \citep{Gonzalez2013,Kravtsov2018,DeMaio2018,Zhang2019,DeMaio2020}. Previous efforts to separate BCG+ICL include using the velocity distribution of stellar particles to distinguish between multiple components \citep{Remus2017}, and attempts to distinguish individual particles by estimating their `boundedness' to the BCG \citep{Cui2014}. \citet{Pillepich2018b} argue that, since the physical origin of the BCG and ICL is the same, the best way to separate them is by means of a fixed spatial aperture, so that masses can be estimated and compared unambiguously across observations and simulations. 
Another physical way to separate the two components is given by the transition radius, that is, the radius at which the diffuse component begins to dominate the stellar mass distribution for BCG+ICL systems in groups and clusters, as recently explored by several authors \citep[e.g.][]{Gonzalez2021,Chen2022,Contini2022,Proctor2023}.

One quantity that is widely used to characterise the ICL is its fractional budget, namely the stellar mass (or light) that is contained in this component out of the total stellar mass (or light) of the cluster. This has been the focus of many studies (e.g. \citealp{Krick2007,Burke2015,Contini2018,MontesTrujillo2018}), but there is no consensus in the specific value for this fraction; it ranges between less than 10 and more than 50 per cent depending on the study. While this depends significantly on the definition of ICL used \citep{Rudick2011,Tang2018,Pillepich2018b,Kluge2021,Ellien2021,Garate-Nunez2023}, it is a clear conclusion that all massive clusters contain a non-negligible ICL component. Given the ambiguity of the mass fraction, other properties are needed to properly characterise the ICL. Some works include the study of the formation of the ICL, by analysing the age and metallicity of its stellar population \citep{MontesTrujillo2018} or its colour gradients and profiles \citep{Morishita2017,DeMaio2018}, or by studying the importance of mergers in ICL formation \citep{Murante2007,Contini2018,Kluge2023}. 

Another consideration that can give insights about the formation of the ICL is the correlation between ICL mass fraction and total cluster mass (noting that these results are also dependent on how the ICL is identified). The observed lack of correlation (e.g. \citealp{Montes2022}) points to a similar ICL formation efficiency across groups and clusters. 
There are still some tensions regarding this result, so that this lack of correlation between ICL fraction and cluster total mass cannot be confirmed (see reviews by \citealp{Contini2021} or \citealp{Montes2022}), but recent results by \citet{Ragusa2023} find no significant trend for a wide halo mass range ($10^{12.5}-10^{15.5} M_\odot$). However, for the dynamical state of the cluster, \citet{Ragusa2023} find a significant trend with ICL fraction, that has also been previously found in other studies, both theoretical \citep{Rudick2006,Contini2014} and observational \citep{DaRocha2008,MontesTrujillo2018,Poliakov2021}. More evolved and hence relaxed clusters tend to have higher ICL fractions, so that the dynamical state of a cluster can be a better indicator of ICL fraction. \citet{Contini2023a} confirm this finding by showing that the main driver of ICL formation is the halo concentration (which is related to both formation time and dynamical state), in a way that more concentrated halos have higher fractions of ICL.

More recent ICL studies have been devoted to the relation between ICL and the dark matter (DM) component of clusters. By studying their density profiles in the IllustrisTNG simulations, \citet{Pillepich2018b} found that the stellar and DM halos of clusters have a similar shape, which was later confirmed by \citet{MontesTrujillo2018} in the Hubble Frontier Fields (HFF) clusters. \citet{MontesTrujillo2019} further observed that there is a tight correlation between the distribution of the diffuse stellar surface density and the surface density of total mass, which effectively means that the ICL can be used to accurately trace the DM distribution. This was theoretically confirmed by \citet{AlonsoAsensio2020} using the Cluster-EAGLE simulations \citep{Barnes2017} and later by \citet{Yoo2022} within a different set of N-body simulations. In contrast, \citet{Sampaio-Santos2021} argue that the question of whether the diffuse light faithfully traces the cluster's radial matter distribution lacks consistent evidence yet. However, they emphasize that the diffuse light remains an excellent indicator of a cluster's total mass. These results are particularly interesting since they provide a way of probing the DM in clusters using only deep imaging observations, and hence highlight the importance of further investigating and understanding this diffuse component of clusters.

Driven by these questions about the properties and formation and evolution of the ICL component, which still remain open, we provide a new study of ICL using cosmological simulations of galaxy clusters. 
In this work, we characterise the intra-cluster light of the clusters in \textsc{The Three Hundred} data set, that consists of regions of radius 15 $\hMpc$ centred on the 324 most massive objects selected from a cosmological dark matter only simulation of side length 1 $\hGpc$. These regions have been re-simulated with two different full physics hydrodynamical codes: the smoothed particle hydrodynamics (SPH) code \textsc{Gadget-X}, and the novel meshless hydrodynamic and gravity solver \textsc{Gizmo} \citep{Hopkins2014,Hopkins2017}. As these simulated regions share the same initial conditions in both data sets, we have the ability to perform code-to-code comparisons on the same clusters. The size of the sample and the mass range included allow for a statistical study of the ICL, analysing also its dependence on different cluster properties. 
This work just forms the first part in a series of papers. Here we simply start with presenting the data set and a first quantitative analysis of it at redshift $z=0$. In a follow-up study (Contreras-Santos, in prep.) we put more focus on the origin of the ICL and direct predictions for observations.

This paper is organised as follows. In \Sec{sec:data} we present \textsc{The Three Hundred} simulations data set, an overview of the galaxy formation models adopted by the simulations, and a description of how galaxies and dark matter halos are identified. In \Sec{sec:identification} we present our method to identify the BCG and ICL components in the simulations, while in \Sec{sec:bcg_icl} we analyse the joint BCG+ICL component of the simulated clusters and compare their stellar mass with observational estimates. In \Sec{sec:icl_frac} we present our resulting ICL mass fraction, and how it depends on cluster total mass and dynamical state. Finally, we study the relation between the ICL and the DM component of clusters in \Sec{sec:icl_dm}, by comparing their density and velocity dispersion profiles. We summarise and discuss the results of this work in \Sec{sec:conclusions}.


\section{The Data} \label{sec:data}

\subsection{The Three Hundred sample}
\textsc{The Three Hundred} project\footnote{\url{https://the300-project.org/}}, first presented in \citet{Cui2018}, is a large international collaboration focused on understanding the formation, evolution and properties of massive galaxy clusters using semi-analytic models and hydrodynamical simulations. The data set consists of 324 massive galaxy clusters with masses above $\sim 8 \times 10^{14} \hMsun$ drawn from the parent dark matter-only simulation MultiDark Planck 2 (MDPL2, \citealp{Klypin2016}), which follows the hierarchical assembly of $3840^3$ dark matter particles in a periodic cosmological volume of 1 $\hGpc$, with a dark matter particle mass resolution of $1.5\times 10^9$ $\hMsun$, and a Plummer equivalent softening of 6.5 $\hkpc$. The simulation adopts a $\Lambda$CDM cosmology with parameter values $\Omega_\mathrm{m}=0.307$, $\Omega_\mathrm{b}=0.048$, $\Omega_\Lambda = 0.693$, $h=0.678$, $\sigma_8 = 0.823$ and $n_s = 0.96$, in agreement with Planck 2015 results \citep{Planck2015}. Dark matter halos in MDPL2 were identified using the \textsc{rockstar} halo finder \citep{Behroozi2013}, from which the 324 most massive halos were selected to constitute the \textsc{The Three Hundred} cluster sample.

In order to perform full hydrodynamical simulations of the selected clusters, it was necessary to use multiple levels of refinement across the entire volume of the simulation using the parallel \textsc{Ginnungagap2} code. This is similar to the so-called zoom technique, but instead of increasing the resolution of a specific region from the parent simulation, the region of interest was kept at the same resolution, and the remainder of the volume was `downgraded' to lower resolution to capture the large scale potential of the box while focusing the computational resources to the region of interest. The `high-resolution' region was delimited by selecting dark matter particles within a sphere of 15 $\hMpc$ centred at the halo of interest at $z=0$, which were then traced back to the initial conditions. These were then split into a dark and a baryonic component with the corresponding $\Omega_\mathrm{m} / \Omega_\mathrm{b}$ ratio, resulting in a dark matter particle mass of $m_{p\mathrm{,dm}}=1.27 \times 10^9$ $\hMsun$, and an initial gas particle mass of $m_{p\mathrm{,gas}}=2.36 \times 10^8$ $\hMsun$. Each one of the selected clusters, together with the regions around them, was resimulated using \textsc{Gadget-X} and \textsc{Gizmo-Simba}, which are described below.

\subsection{Hydrodynamical models}

\subsubsection{Gadget-X}
\textsc{Gadget-X} is based on \textsc{Gadget-3}, a tree particle mesh (TreePM) smoothed particle hydrodynamics (SPH) code\footnote{See \citet{Springel2005} for a detailed description of the last public version \textsc{Gadget-2}}, but it includes major improvements over the standard developer version. Hydrodynamic equations are solved using a density-entropy formulation of SPH \citep{SpringelHernquist2002} using a Wendland $C^4$ kernel \citep{Wendland1995}. 
Gas cooling in the simulation is implemented following \citet{Wiersma2009}, while the adopted star formation model is the same as the one described in \citet{Tornatore2007}, and follows the star formation algorithm presented in \citet{SpringelHernquist2003}. Each one of the created star particles represents a simple stellar population (SSP, i.e. all stars have the same metallicity) with the number density following a \citet{Chabrier2003} initial mass function (IMF). For each of these SSPs, stellar lifetimes are considered following \citet{PadovaniMatteucci1993}, to account for the appropriate mass-loss and metal release timescale of asymptotic giant branch (AGB) stars, supernovae Ia (SN Ia) and corecollapse supernovae (SN II) individually. These events restore mass and enrich the chemical composition of surrounding gas particles with metal yields of AGB stars are taken from \citet{vandenHoekGroenewegen1997}, for SNe Ia from \citet{Thielemann2003}, and for SN II from \citet{WoosleyWeaver1995}; being the latter the only contributor to kinetic stellar feedback, implemented as in \citet{SpringelHernquist2003}.

The adopted black hole (BH) growth and active galactic nuclei (AGN) follows \citet{Steinborn2015}, which is an extended model of \citet{Springel2005}. Black holes in the simulation are implemented as collisionless sink particles that are created at the centre of friends-of-friends (FoF) objects whose mass exceeds $2.5 \times 10^{10}$ $\hMsun$ and do not host a black hole, with a seed mass of $M_\bullet = 5 \times 10^6$ $\hMsun$. The black holes mass accretion rate, $\dot{M}_\bullet$, follows a Bondi-Hoyle-Lyttleton \citep{HoyleLyttleton1939,BondiHoyle1944,Bondi1952} scheme, capped at the Eddington limit, multiplied by a factor of $\alpha = 100$\footnote{\citet{Steinborn2015} proposed a two-mode accretion of hot and cold gas, each with $\alpha_\mathrm{hot}$ and $\alpha_\mathrm{cold}$, respectively. However, for this implementation only the cold-gas accretion is used, or equivalently a value of $\alpha_\mathrm{hot}=0$ is adopted.}. The feedback energy input into the surrounding is implemented as purely thermal.

This implementation of \textsc{Gadget-X} has been used to study properties of galaxy clusters such as the cool core/non-cool core cluster dichotomy \citep{Rasia2015}; to quantify the abundance and spatial distribution of neutral hydrogen \citep{VillaescusaNavarro2016}; to study the hot gas pressure profiles, gas clumping and Sunyaev-Zeldovich (SZ) scaling relations \citep{Planelles2017}; and the galaxy cluster X-ray scaling relations \citep{Truong2018}. As part of \textsc{The Three Hundred} project, \textsc{Gadget-X} simulated clusters have been used to study galaxy and cluster population relations \citep{Cui2018}, the self-similar evolution of galaxy cluster density profiles \citep{Mostoghiu2019} or the radial and galaxy-halo alignment of dark matter subhalos and satellite galaxies \citep{Knebe2020} among others.

\subsubsection{Gizmo-Simba}

The \textsc{Gizmo-Simba} runs of \textsc{The Three Hundred} are performed with
the \textsc{Gizmo} code \citep{Hopkins2015}, with the state-of-the-art galaxy formation subgrid models following the \textsc{Simba} simulation. 
\textsc{Simba}, first introduced in \citet{Dave2019}, is a set of cosmological hydrodynamical simulations run with a modified version of the gravity and hydrodynamics solver \textsc{Gizmo} \citep{Hopkins2015,Hopkins2017}, with improved star formation and feedback subgrid models from its predecessor \textsc{Mufasa} \citep{Dave2016}, and including recipes for black hole growth and its corresponding feedback. Although \textsc{Simba} on its own is a separate set of simulations, the same galaxy formation model was used to run the 324 selected regions of  \textsc{The Three Hundred} project. We therefore refer to the overall code-model implementation as \textsc{Gizmo-Simba}. Below we briefly describe the relevant details of the code and subgrid models.

\textsc{Gizmo} is a massively parallel gravity and hydrodynamics solver first presented in \citep{Hopkins2015} and last updated in \citep{Hopkins2017}. The skeleton of the code is largely based upon \textsc{Gadget-3} \citep{Springel2005}, and while SPH is preserved within the code (amidst some modifications), its main feature is the implementation of so-called Meshless Finite Mass and Volume (MFM and MFV, respectively) methods to solve hydrodynamics. This can be oversimplified as a `mixture' of SPH and moving-mesh approaches that conserves mass, energy, momentum and angular momentum, that is capable of accurately capturing shocks and fluid mixing instabilities, without the need for artificial viscosity. 

Photoionisation heating and radiative cooling of gas in \textsc{Simba} is computed using the \textsc{Grackle-3.1} library \citep{Smith2017}. 
Star formation is modelled using an $\mathrm{H_2}$-based approach where the $\mathrm{H_2}$ fraction is computed following \citet{KrumholzGnedin2011}, accounting as well for variations in numerical resolution, as described in \citet{Dave2016}. In its standard implementation, a star-forming gas particle produces a single generation of `stars', for which its entire mass is converted into a single stellar particle. Stellar feedback accounts for the contribution from AGB stars, and from supernovae type Ia and II. AGB stars and Type Ia SNe produce galactic winds using \citet{Dave2016} two-phase model with a fraction of ejected hot gas to be 30 per cent.

Black holes in the simulations are seeded in FoF galaxies that reach stellar mass $M_* \gtrapprox 10^{10.5}\Msun$ and do not already contain a BH particle by converting the star particle closest to the centre of mass into a BH particle. Black hole particles are repositioned to the minimum of the gravitational potential of their host FoF galaxy if it is located within 4 $R_0$, where $R_0$ is the size of the black hole kernel used for accretion. Black hole particles are allowed to merge if two of them are located within $R_0$ and their relative velocity is smaller than three times their mutual escape velocity. \textsc{Simba} implements a novel black hole mass accretion model using a two-mode approach for hot and cold gas accretion. In this model, the accretion of hot gas follows the standard Bondi-Hoyle-Lyttleton \citep{Bondi1952} spherical accretion, while the accretion of cold gas is modelled using the torque-limited approach from \citet{HopkinsQuataert2011} and its numerical implementation from \citet{AnglesAlcazar2017}. 

\textsc{Simba} accounts for two different feedback mechanisms associated to black holes: kinetic outflows, which arise from a `radiative' mode at low accretion rates ($\dot{M}_\mathrm{BH}$) and from `jet' modes at high accretion rates; and feedback from X-ray sources. The addition of feedback from accretion disk X-rays is a novel feature of \textsc{Simba} and, although it does not contribute largely to the overall feedback energy release, is key to quenching the most massive objects \citep{Dave2019}. Note that the \textsc{Gizmo-Simba} run of \textsc{The Three Hundred} clusters uses recalibrated parameters for the baryon model due to the different simulation resolutions \citep[see][for details]{Cui2022}.

\subsection{Halo catalogues}
We use Amiga Halo Finder (AHF\footnote{\url{http://popia.ft.uam.es/AHF}}, \citealp{Gill2014,KnollmannKnebe2009}) to identify dark matter halos in the simulations. We use the same halo catalogues as previous studies from \textsc{The Three Hundred} collaboration for consistency and to take advantage of the estimated properties and derived formation histories of the halos in other studies (e.g. \citealp{Cui2018,Mostoghiu2019}). Halos are identified by locating density peaks in the simulation volume in a similar fashion as adaptive mesh refinement (AMR) techniques, in which space is iteratively divided when the total matter density within a cell exceeds a given threshold to constrain the location of overdense regions. Once these density peaks are found, particles of all species (dark matter, stars, gas and black holes) are grouped into gravitationally bound halos. The halos are then spherical regions composed by at least 20 particles with a total matter density of $\Delta$ times the redshift-dependent critical density of the Universe, $\rho_\mathrm{crit}$, with a mass of $M_\mathrm{halo} = \Delta \cdot \rho_\mathrm{crit}(z) \cdot R_\mathrm{halo}^3 \cdot 4\pi /3$, where $R_\mathrm{halo}$ is the corresponding spherical overdensity radius. Two catalogues have been produced with enclosing overdensities of $\Delta = \{ 200, 500 \}$; physical quantities corresponding to these will be denoted by the subscript 200 and 500, respectively, whenever necessary.
Regarding the substructures, AHF defines subhalos as halos that lie within $R_\mathrm{halo}$ of a more massive halo, which is called the host halo. The mass of this host halo then includes the masses of all the subhalos contained within it.

\section{Identification of BCG and ICL} \label{sec:identification}

In general, the BCG is defined to be the brightest galaxy in the cluster, while the ICL is comprised by the stars that are not bound to any galaxy (central or satellite) but to the potential of the cluster itself. However, these definitions are not trivial to implement in observations of galaxy clusters, and so identifying the different components is a challenging task. This way, separating the outer regions of the BCG from the ICL is still an ill-defined problem \citep{Gonzalez2005,Krick2007,Jimenez-Teja2016}. Moreover, the low surface brightness of the ICL \citep{Mihos2005,Zibetti2005,Rudick2010} and its contamination by foreground and background galaxies make it even more complicated. 
In simulations, this problem is simplified by the availability of full kinematic information of the baryonic and dark matter particles. However, it is still not trivial to separate the gravitational potential of the host cluster from that of the individual galaxies -- and thus the BCG and the ICL. For this reason, different works can follow different approaches, and it is hence important to clarify how this is done. In this section we explain how we define and find the BCG and the ICL in the galaxy clusters of \textsc{The Three Hundred} simulations. 

 \begin{figure*}
  \centering
  \includegraphics[width=12.5cm]{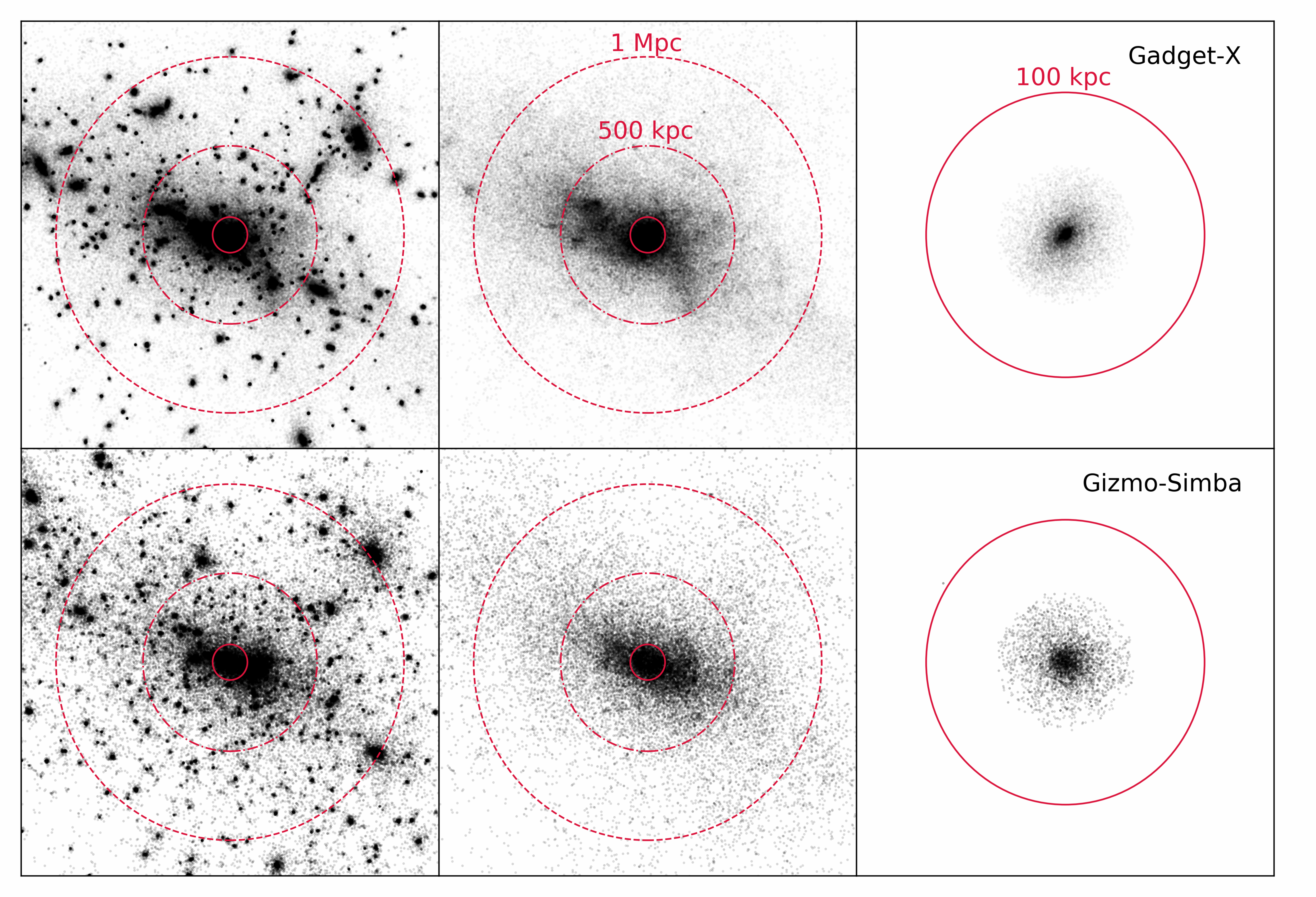}
  \caption{For one of the 324 clusters (cluster 6), 2D projection of all the stellar particles along one axis that belong to, from left to right, the total stellar mass within $R_{500}$, the ICL and the BCG alone.  The same colour intensity is used for all the visualizations with darker tones showing higher densities; concentric circles have radii of 100 kpc (inner), 500 kpc (middle), and 1 Mpc (outer). For the rightmost plots only the inner circles (100 kpc radii) are shown.}
\label{fig:icl_example}
\end{figure*}

\subsection{Defining the BCG} \label{sec:identification:bcg}
Since the edge of a BCG is not clearly delimited, we determine it by selecting all the stars inside a sphere of certain radius, using the halo centre as the origin. This way of defining the BCG, based on a fixed physical radius, has often been adopted in simulations. For instance, \citet{McCarthy2010} used a radius of 30 $\hkpc$, whereas in \citet{Ragone-Figueroa2018} three different radii were used too (30 kpc, 50 kpc and $0.1R_{500}$). \citet{Pillepich2018b} further consider that this rather arbitrary separation between BCG and ICL is validated by theoretical arguments, since both components are formed by smooth stellar mass accretion and mergers, although in different relative amounts. They use a separation of 30 kpc, and consider also the effect of changing it to 100 kpc.
From a more observational approach, \citet{Kravtsov2018} advocate the use of stellar masses defined this way for comparison with observations, and they provide values for nine BCGs using several different radii. Besides, \citet{Stott2010} showed that, using a 50 kpc aperture radius, BCG luminosities can be recovered with less than 5 per cent difference from those obtained in some observational analyses. 

Hence, using a fixed aperture allows us to easily identify in the simulation all the stellar particles that are in this central region. As we have seen, different values for this aperture are used throughout the literature. In a previous work with the same simulations, we obtained the BCG for three different fixed apertures: 30, 50 and 70 kpc \citep{Contreras-Santos2022a}. In this work, for clarity and simplicity, we will focus on the 50 kpc aperture and, unless specified otherwise, show the results only for this aperture.  Please refer to Sec.~\ref{sec:id_aperturesize} and the Appendix~\ref{appendix:aperture} for results based upon the other apertures.

Despite our choice, there are some caveats of this BCG definition that need to be mentioned. First, we adopt a 3D cut, this is, we consider the stars outside the central sphere on the foreground and background to belong to the ICL, which is not fully reproducible in observations. In this regard, \citet{Cui2022} verified that using the projected 2D cut increases the BCG masses by less than 20 per cent. Additionally, this definition can introduce some bias if applied to a sample with a wider mass range, where the sizes of the central galaxies can change significantly from group to cluster scales. Alternative methods that look for the transition radius from BCG to ICL \citep[e.g.][]{Contini2022,Chen2022} can be valuable tools in overcoming these challenges, while offering a better physical understanding of the situation. However, for our study, limited to high masses, a fixed aperture has proven adequate to identify the BCG without introducing any significant bias. 


\subsection{Defining the ICL}
Once the BCG is defined and the particles that assemble it are selected, we can easily find the ICL in our simulations. In order to do this, we select all the stellar particles that are bound to the central cluster, but do neither belong to the BCG nor to any subhalo. This separation of particles can be easily done using the AHF catalogues, which give the IDs of all the particles that belong to each of the halos and subhalos, for the simulations run with both \textsc{Gadget-X} and \textsc{Gizmo-Simba} codes. In other words, from all the stellar particles in the halo, we first remove the BCG (whose definition is solely based on a fixed aperture of 50 kpc, not dependent on the halo finder) and then we remove the particles bound to substructures (as indicated by AHF). The remaining particles are the ICL of the cluster. Note that in special situations, such as major mergers, a large secondary halo can bring its own ICL. In this case, as long as the secondary halo is identified as a different object from the main one, all its particles are removed for the counting of the main halo ICL.

In \Fig{fig:icl_example}, for one cluster selected as example, we show a 2D projection of all the stellar particles along one axis of our simulations for, from left to right, total stellar mass within $R_{500}$, the ICL and the BCG alone. Top row is for \textsc{Gadget-X} and bottom row for \textsc{Gizmo-Simba}. The same colour intensity is used for all the visualizations, with darker tones showing higher densities. In the second column, which corresponds to the ICL, we can see how the substructures in the cluster, that were present in the first column, have been removed, leaving the ICL as a diffuse component. 
Since we use a 3D aperture to extract the BCG, we still see particles in the innermost 50 kpc when projecting them along one axis, but they are in fact outside the central sphere, either in the front or in the background. 
We want to note that particles in the outskirts of galaxies might not be identified to belong to the galaxy, and thus still be part of the ICL. Due to projection effects, these would appear in \Fig{fig:icl_example} as small clumps, but would be in fact empty shells, not representing a significant part of the ICL of the cluster. 
The last column in \Fig{fig:icl_example} shows a zoom-in to the BCG, which, as described before, is selected as all the stellar particles within 50 kpc of the halo centre. The red circle in that panel depicts again a sphere of radius 100 kpc. We note the BCG's stellar particle distribution is concentrated towards the centre, showing a great reduction towards the outer parts. We are therefore confident that the BCG has properly been excised and separated from the ICL (but see also the discussion in \Sec{sec:id_aperturesize}). 

Finally, comparing the top and bottom rows in \Fig{fig:icl_example}, we can see that in general the same structures are generated by both codes, with differences being found only on smaller scales. We can also see that in \textsc{Gadget-X} there are more particles, making the maps look smoother, while by the colour intensity we can see that the mass is very similar in both simulations. This points to different stellar particle resolution between the two simulations, due to their different implementations of star formation. We will discuss this in more detail in the following section.

 \begin{figure}
  \hspace*{-0.35cm}
  \includegraphics[width=8.3cm]{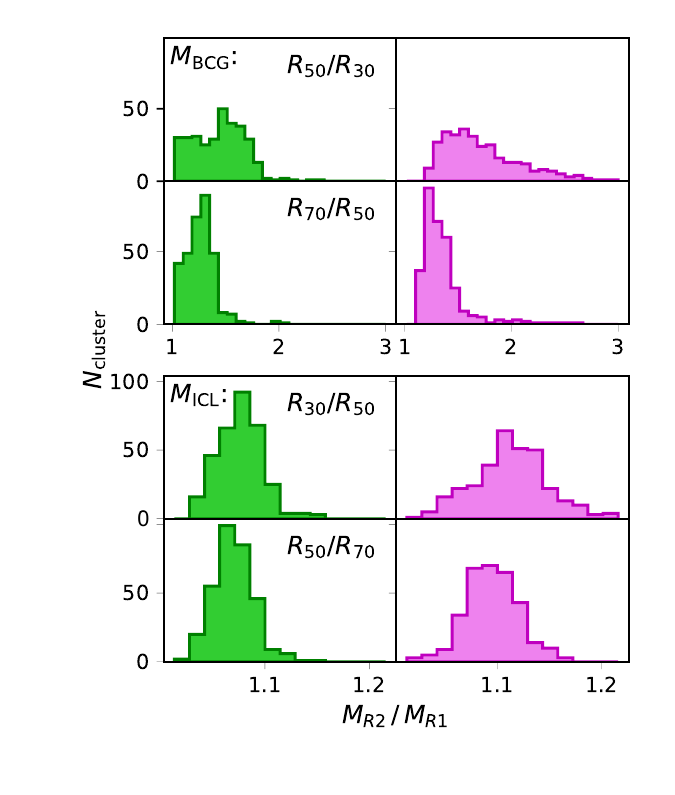}
  \vspace*{-0.6cm}
  \caption{Dependence of BCG and ICL mass on aperture size. Top: distribution of the ratio between the BCG mass of the \textsc{The Three Hundred} simulations obtained for apertures: 30 and 50 kpc (first row), and 50 and 70 kpc (second row). The left column (in green) is for \textsc{Gadget-X}, while the right one (in magenta) is for \textsc{Gizmo-Simba}. Bottom: same for ICL mass. Since ICL mass decreases with increased aperture, the ratio is inverted now to keep the values above 1, but the interpretation remains the same.}
\label{fig:apertures}
\end{figure}

\subsection{Dependence on aperture size} \label{sec:id_aperturesize}
Since our BCG definition involves a `free' parameter, namely the size of the fixed aperture, the resulting ICL also depends on the value used for it. Although we have chosen a value of 50 kpc for this parameter, before continuing with our results we believe it is relevant to show how sensitive the resulting BCG and ICL are to the size of the aperture.

In \Fig{fig:apertures} we compare the masses of the two components, as computed using different apertures. The upper $2 \times 2$ subplot compares the BCG mass, the left column (in green) being the results for \textsc{Gadget-X} and the right one (in magenta) for \textsc{Gizmo-Simba}. The top row shows a histogram of the ratio $M_\mathrm{BCG,50}/M_\mathrm{BCG,30}$, where we can see that the distribution is wide, reaching even more than 2, which means that the BCG mass is being doubled when changing the aperture from 30 to 50 kpc. In the second row the ratio shown is $M_\mathrm{BCG,70}/M_\mathrm{BCG,50}$, and in this case we see that the values are significantly smaller, the median growth being around 30 per cent. This indicates that the difference between the BCGs computed for 70 and 50 kpc apertures is not as relevant as in the first row, which suggests that increasing (moderately) the aperture beyond 70 kpc does not make such a significant difference in the BCG mass.

The lower $2 \times 2$ subplot in \Fig{fig:apertures} shows this same comparison but for the ICL mass. Since the BCG mass increases for increased aperture, the ICL mass is reduced. However, to keep the histogram values above 1, we now plot the inverse ratios. The meaning of the plots remains the same, the further the values are from 1 the more the mass is changing from one aperture to the other. We see in this case that the values are much closer to 1 than for the BCG masses; since the ICL extends quite far from the halo centre, the choice of a fixed aperture between 30 and 70 kpc does not affect that much the final mass of this component.


In conclusion, \Fig{fig:apertures} shows that, while for the BCG the size of the aperture selected can have an important effect on the resulting BCG mass, the ICL mass is not very much dependent on this value. Since the main focus of the paper is on the ICL component of clusters, we believe these differences should not be discussed in the main text of the paper. However, we include an appendix at the end of this work where we show how the main results of the paper, derived in the following sections, are affected by the size of the aperture, for which we use 30, 50 and 70 kpc. From now on in this paper -- and unless specified otherwise -- we will use a fixed aperture of 50 kpc to define the BCG. For the results for 30 and 70 kpc apertures, we refer the reader to \App{appendix:aperture}. We note that, although both the BCG and ICL masses are affected by the aperture size, by definition the joint component BGC+ICL is independent of it. We will start the next section by studying this component and comparing our findings to those of previous observations.



\section{BCG+ICL} \label{sec:bcg_icl}

In this section we investigate the stellar mass content and distribution of the BCG+ICL, $M_\mathrm{*BCG+ICL}$, as a function of cluster halo mass $M_{500}$. We start with comparing against observations for which the BCG+ICL is available to understand similarities and differences intrinsic to the simulations themselves and the implications these may have on other results.

\begin{figure}
  \centering
  \includegraphics[width=8.3cm]{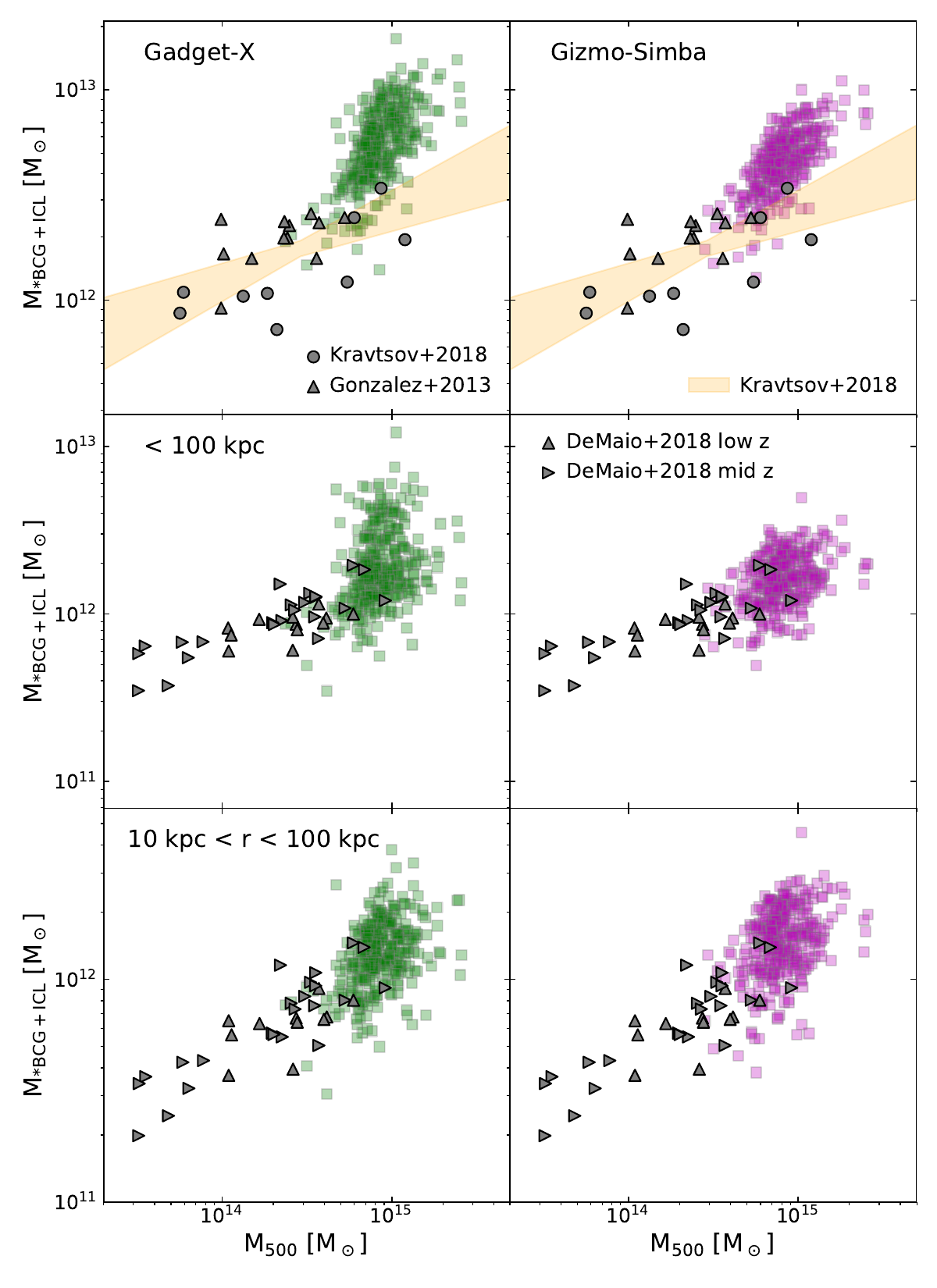}
  \caption{\textsc{Gadget-X} (left) and \textsc{Gizmo-Simba} (right) BCG+ICL stellar mass within different spherical regions as a function of total halo mass. First row shows the total BCG+ICL stellar mass, squares denote individual measuremente of each of the 324 main clusters, grey symbols show data from \citet{Gonzalez2013} and \citet{Kravtsov2018}, as labelled; the shaded region shows the best-fit from \citet{Kravtsov2018}. Second row shows the total BCG+ICL mass within 100 kpc aperture and the third one shows the BCG+ICL stellar mass contained between 10 and 100 kpc apertures. Grey symbols show observational estimations from \citet{DeMaio2020}.}
\label{fig:bcgicl_obs}
\end{figure}

In \Fig{fig:bcgicl_obs} we show the stellar mass of the BCG+ICL as a function of halo mass, $M_{500}$. The first row shows the BCG+ICL total stellar mass, the second row shows the total mass inside an aperture of 100 kpc and the bottom row shows the mass content between 10 and 100 kpc. Individual estimates from the simulated clusters are shown as squares, and observational estimates are shown as grey symbols in each panel, as labelled; we note that although multiple groups and clusters may be resolved within the re-simulated regions, these plots only show the main 324 cluster sample. 
Regarding the total BCG+ICL mass (top row), both simulations are able to reproduce observational estimates from \citet{Gonzalez2013} and \citet{Kravtsov2018} at $M_{500} < 10^{14.5} M_\odot$. However, at larger masses both simulations overpredict observational estimates and best fit line of \citet{Kravtsov2018}, with \textsc{Gadget-X} showing a larger spread than \textsc{Gizmo-Simba} on the $M_\mathrm{*BCG+ICL}$ reaching $\sim$ 1 dex at $M_{500} \simeq 10^{15} M_\odot$. Note that the \textit{total} BCG+ICL mass in \citet{Kravtsov2018} corresponds to the mass within $R_\mathrm{out}$, defined as the radius where the error in fitted Sérsic profiles reaches 33 per cent due to background uncertainty. Furthermore, since observations are limited by the sensitivity of the instruments to detect the faint intra-cluster light and are prone to projection effects, our comparison is not fully consistent. Using 3D apertures that include all ICL particles, some excess mass in the simulations can be expected.

The BCG+ICL mass within a 100 kpc aperture (second row in \Fig{fig:bcgicl_obs}) shows different mass contents for \textsc{Gadget-X} and \textsc{Gizmo-Simba}. This difference is only seen in the second row, and not in the third one, which shows the mass between 10 and 100 kpc apertures. This means that the difference between the two simulations is present in the innermost parts of the BCG+ICL ensemble. Compared to observational estimates from \citet{DeMaio2020}, for apertures of 100 kpc (second row), \textsc{Gizmo-Simba} shows a notable agreement with observations, while \textsc{Gadget-X} shows bigger scatter, with up to $\sim$ 2 dex spread. Lastly, regarding the mass content in between 10 and 100 kpc, both simulations show a good agreement with \citet{DeMaio2020} within the halo mass range where observations and simulations overlap. Although the trend is well followed at $M_{500} > 10^{15} M_\odot$, this does not seem to hold for masses below the range contained within the simulations. From this last row we can infer that the observed spread in the 100 kpc aperture estimate comes from the stellar mass contained within the innermost 10 kpc.

\begin{figure}
  \centering
  \includegraphics[width=8.2cm]{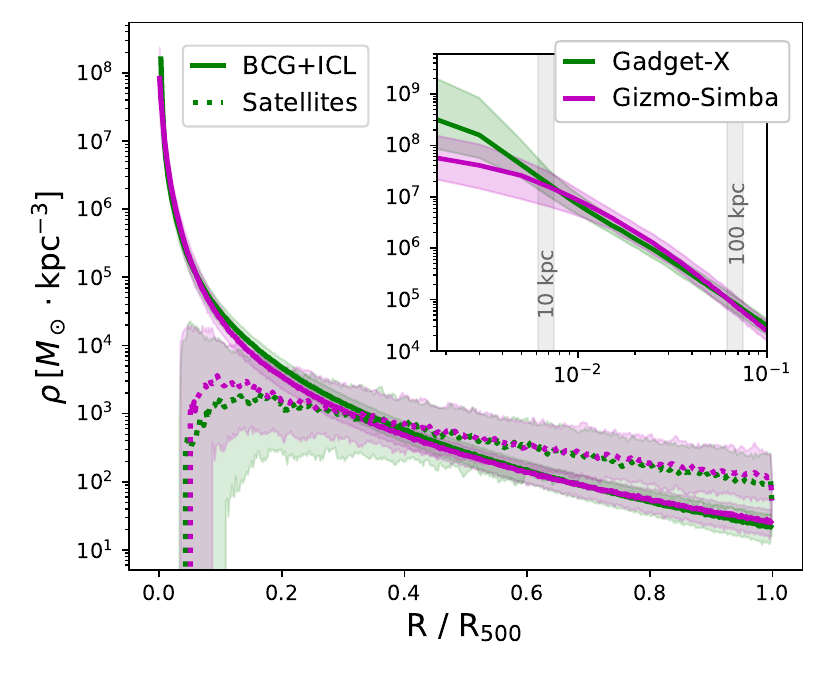}
  \caption{Density profile of the BCG+ICL component (solid) and satellite galaxies (dotted) for the 324 main clusters of \textsc{Gadget-X} (green) and \textsc{Gizmo-Simba} (magenta) scaled by cluster spherical overdensity radius, $R_{500}$. Lines shows the median of each bin, and shaded regions delimit the 16th and 84th percentile of each bin. Inset shows only the BCG+ICL component in logarithmic scale in the X-axis, to focus on the inner part of the profile. Here, the vertical shaded regions show for reference the 16th and 84th percentile of the location of 10 and 100 kpc, as labelled.}
\label{fig:bcgicl_profile}
\end{figure}

We further investigate the BCG+ICL component in \Fig{fig:bcgicl_profile}, which shows the volumetric-density profile of this component, scaled by cluster radius, $R_{500}$, stacked for the 324 clusters in \textsc{The Three Hundred} sample. The solid lines show the median density value at a given bin, while the shaded regions indicate the 16th and 84th percentiles, with the different simulations coloured as labelled. For comparison, we also include the density profile of the satellite galaxies (i.e. including only stellar mass) as dotted lines. The inset in \Fig{fig:bcgicl_profile} shows only the BCG+ICL component, in logarithmic scale in the X-axis, zooming in for the innermost region of the profile, up to 0.1$R_{500}$ (showing the location of 10 and 100 kpc as vertical regions for reference).

The profiles show some important features. First, the profiles of both samples display very similar amplitudes and slopes between 0.01 and 0.8 $R_{500}$ with a 1 $\sigma$ scatter in the density of 0.3 dex, suggesting that the stellar distributions are self-similar at least within the mass range of the sample. This not only confirms the agreement in the mass contained in between 10 and 100 kpc between the samples, but also suggests that we should expect similar results for the annulus up to $\sim 0.8 R_{500}$, after which the value of the density becomes significantly low, showing the `edge' of the BCG+ICL component. The remarkable agreement of $M_\mathrm{*BCG+ICL}$ within $10 < R < 100$ kpc (see inset) demonstrates that the overall physical processes responsible for the formation of the BCG and the ICL around it are well modelled by both simulations. Regarding the comparison with satellite galaxies, we can see that, in median, satellites start to dominate over the diffuse component at $\sim 0.35 R_{500}$, this being slightly closer to the centre for \textsc{Gizmo-Simba}. However, the satellites' profiles show a wide scatter, so that this threshold can range from $\sim 0.2$ to $0.6 R_{500}$ for the different clusters.

Focusing now on the inset in \Fig{fig:bcgicl_profile}, we can see that the density profiles of the two simulations' data sets differ the most within the innermost radii, $R < 0.01 R_{500}$, where \textsc{Gadget-X} clusters show an upturn in the density profile, displaying a slope of $\sim -2$, while \textsc{Gizmo-Simba} start to flatten towards the centre, with a slope of -0.9. The density difference between the two data sets' medians within this region is $\gtrsim 0.5$ dex, showing that \textsc{Gadget-X} produces denser and more massive centres than \textsc{Gizmo-Simba}, which explains the difference observed in the 100 kpc aperture measurements of \Fig{fig:bcgicl_obs}, where \textsc{Gadget-X} displays a large scatter towards higher stellar masses at fixed halo mass.
Moreover, the large spread observed for both simulations in $M_\mathrm{*BCG+ICL}$ within 10 kpc apertures is tightly connected to the numerical resolution of the simulations and their specific implementation of the star formation. On one side, \textsc{Gadget-X} and \textsc{Gizmo-Simba} were run with softening lengths of $6.5\hkpc$ and $5 \hkpc$, respectively, which means that the kinematics within 10 kpc apertures are very close to the scales that can be considered as well resolved. Some of the differences between the two simulations are likely coming from the different stellar particle resolutions, recalling that the implementation of \textsc{Gadget-X} allows gas elements to produce multiple generations of stars, while for \textsc{Gizmo-Simba} the entire mass of a gas particle is converted into a single star. This can make the stars produced by the two simulations to differ significantly in mass, although the initial dark matter and gas mass resolution are the same between both simulations. A direct consequence of distributing the same mass amongst a fewer number of particles is that the system becomes more collisional and therefore the dynamics in the simulation are more prone to be affected by numerical scattering (e.g. \citealp{BinneyKnebe2002,Ludlow2019}), as the mass ratio between dark matter particles and star particles is $\sim 10$ for \textsc{Gizmo-Simba} and $\sim 40$ for \textsc{Gadget-X}.

\section{ICL mass fraction} \label{sec:icl_frac}
In the previous sections we focused on explaining how we define both the BCG and ICL of clusters and presenting results for the joint component BCG+ICL. In this section our aim is to characterise the ICL in \textsc{The Three Hundred} clusters, investigating how much of the total stellar mass of the cluster can be found in this component and how this ratio might depend on different cluster properties.

\subsection{Mass fraction of ICL}

\begin{figure}
  \centering
  \includegraphics[width=7cm]{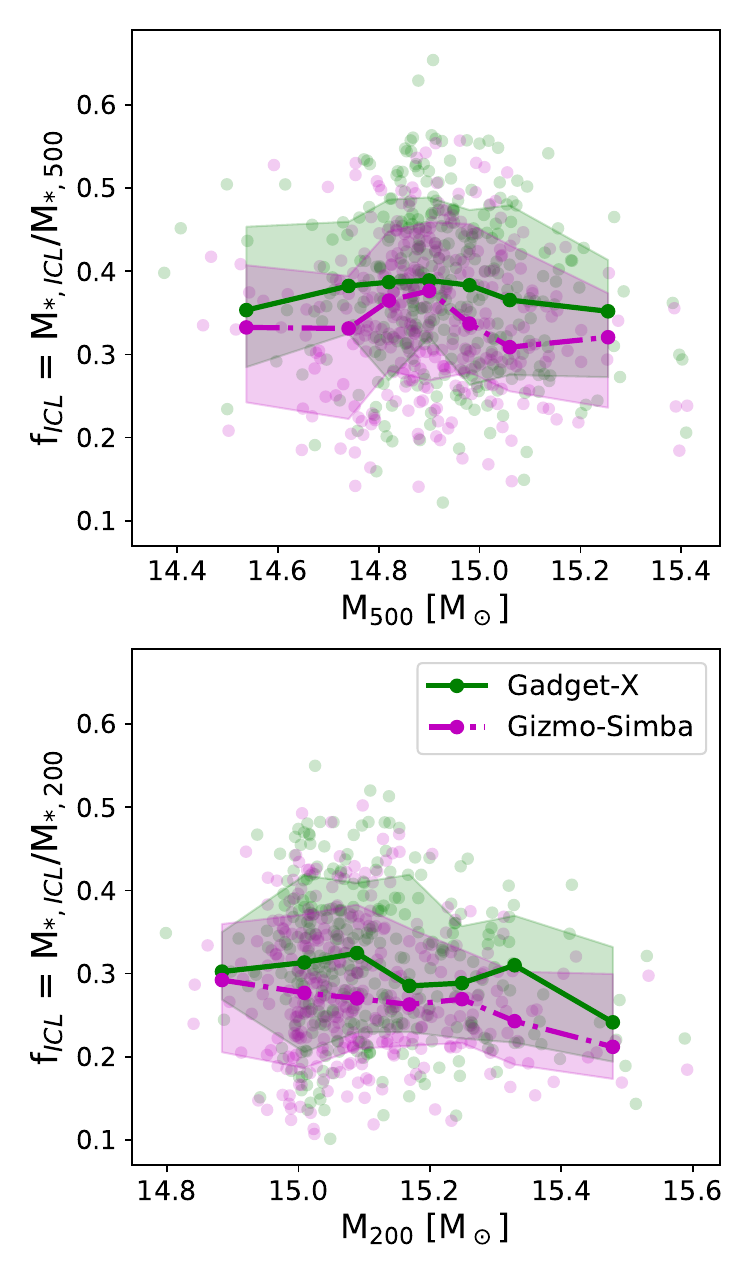}
  \caption{ICL mass fraction as a function of cluster mass, considering overdensities of 500 (top) and 200 (bottom). The lines show the median values and the shaded regions the 16th-84th percentiles. Results are shown for both \textsc{Gadget-X} (green solid line) and \textsc{Gizmo-Simba} (magenta dash-dotted line). Note that the scale is the same for both plots in the Y-axis, but different for the X-axis.}
\label{fig:massfraction}
\end{figure}

We first compute the mass fraction of the ICL within the cluster, dividing the stellar mass found in the ICL as defined in \Sec{sec:identification} by the total stellar mass within the considered overdensity ($M_{*,200}$ or $M_{*,500}$). In \Fig{fig:massfraction} we show this fraction as a function of the cluster's halo mass, in the top panel using $M_{500}$ and in the bottom $M_{200}$. The dots show the individual values for all 324 regions, while the lines (solid for \textsc{Gadget-X} and dash-dotted for \textsc{Gizmo-Simba}) indicate the median values for the different mass bins. The corresponding shaded regions depict the 16th-84th percentiles, in green for \textsc{Gadget-X} and magenta for \textsc{Gizmo-Simba}. 

The main feature that stands out in \Fig{fig:massfraction} is that the ICL mass fraction shows no apparent dependence on cluster mass for our considered mass range. The specific value is different between $R_{200}$ and $R_{500}$, with the fraction being higher for the innermost overdensity of $R_{500}$. In this case it has an approximately constant value of $f_\mathrm{ICL}\sim0.36$, while for $R_{200}$ the value is around $0.29$. Another important result to highlight here is that both hydrodynamical simulations \textsc{Gadget-X} and \textsc{Gizmo-Simba} show the same results, with no significant difference between them.
This is in agreement with recent studies of ICL at low redshifts, that have also shown no dependency of the ICL mass fraction with cluster mass \citep{Montes2022,Ragusa2023}, for our considered mass range. 
Theoretical studies such as \citet{Contini2023a} or \citet{Proctor2023} further support this scenario, with no discernible correlation for our cluster mass range.

\begin{figure}
  \centering
  \includegraphics[width=7cm]{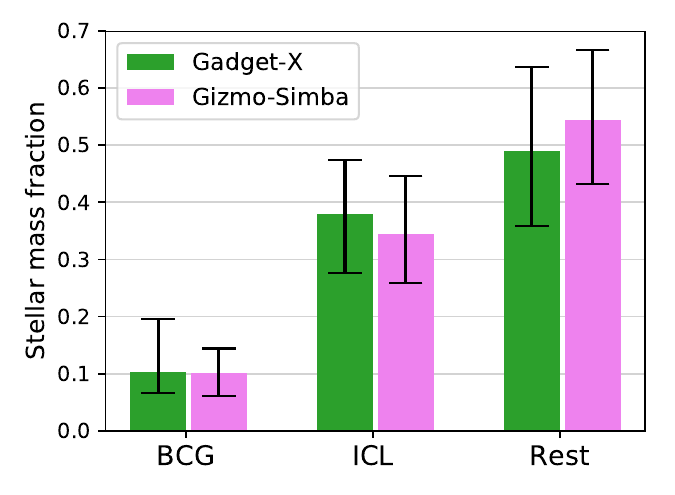}
  \caption{Stellar mass fraction for the BCG, ICL and rest of stars. The bar shows the median value for the 324 clusters, while the error bars indicate 16th-84th percentiles. The fraction is computed dividing the stellar mass content in each of the components by the total stellar mass inside $R_{500}$, $M_{*,500}$}
\label{fig:fractions_barplot}
\end{figure}

We also show in \Fig{fig:fractions_barplot} the stellar mass fraction in the other two components: the BCG and the rest of the galaxies in the cluster, now focusing only on the overdensity 500. In this case we show the median value for the 324 clusters, together with the 16th-84th percentiles. The green bars are for \textsc{Gadget-X}, while the magenta ones are for \textsc{Gizmo-Simba}. We see that the results are similar for the two simulations, with the difference that there is higher scatter in \textsc{Gadget-X}, especially for the BCG, where the median is similar for both but the fraction in \textsc{Gadget-X} can be as high as $\sim 20$ per cent, while it does not reach 15 per cent in \textsc{Gizmo-Simba}. This is the same effect we saw in the inner region of the density profiles in \Fig{fig:bcgicl_profile}, due to the differences in numerical resolution and implementation of star formation between the two codes. Regarding the ICL, we see that the fraction is between 30 and 50 per cent in both cases (this result is the same as \Fig{fig:massfraction}, without binning by cluster mass). Finally, for the rest of the galaxies we also see a higher scatter in \textsc{Gadget-X}, with the fraction ranging from 36 to 64 per cent; while for \textsc{Gizmo-Simba} the range is between 43 and 67 per cent. For overdensity 200 we are not showing the results here, but the conclusions are very similar, with a smaller ICL fraction (as we could see in \Fig{fig:massfraction}), and hence a larger contribution from the satellite galaxies.

In general, we can conclude from \Fig{fig:fractions_barplot} that the BCG mass fraction is below 15 and 20 per cent for \textsc{Gizmo-Simba} and \textsc{Gadget-X}, respectively. Compared with the ICL fraction, which is between 30 and 50 per cent, this means that, in terms of mass, the BCG+ICL component is dominated by the ICL when considering the whole range up to $R_{500}$ (or $R_{200}$). The whole cluster stellar content, however, is in general dominated by the satellite galaxies, which can account for more than half the stellar mass of the cluster. This is a high percentage when compared to the situation for the total mass of clusters, that is including not only stars but also DM and gas. In this case, the substructures account for a median of only 10 per cent for our \textsc{The Three Hundred} sample.

How do our results fare against other studies previously presented in the literature, either observationally or theoretically? The ICL mass fraction is a property that has been investigated by several studies, but it shows a great dispersion in the reported values. For 6 clusters at redshift $0.3 < z < 0.6$ observed by the Hubble Frontier Fields (HFF) survey, \citet{MontesTrujillo2018} measured an ICL fraction of $\sim 7$ per cent for $R<R_{500}$, in agreement with previous measurements of the same objects by \citet{Morishita2017}. Around redshift 0.2–0.3, \citet{Burke2015} found the ICL fraction to be around 20 per cent in the CLASH survey data, while \citet{Krick2007} estimated the fraction to be 6–20 per cent in 10 nearby clusters. Using DES observations, \citet{Zhang2019} found the contribution of BCG+ICL to be 44 $\pm$ 17 per cent of the total cluster stellar luminosity within 1 Mpc, for a sample of 300 clusters at redshift in the range 0.2 - 0.3. For lower redshifts, \citet{Gonzalez2007,Gonzalez2013} estimated the BCG+ICL component to make up 33 per cent of the total cluster stellar mass within $R_{200}$, or 25 to 55 per cent within $R_{500}$. In the Local Universe ($z < 0.05$), \citet{Ragusa2023} found that the ICL fraction ranges from 5 to 50 per cent using VEGAS images. On the theoretical side, \citet{Rudick2011} used a set of numerical $N$-body simulations to find that the fraction of the total cluster's luminosity that is in the ICL ranges from 9 to 36 per cent at $z=0$, with a strong dependence on the ICL definition used. The semianalytical models by \citet{Contini2014,Contini2018} predict an ICL fraction that varies between 20 and 40 per cent depending on the particular implementation adopted. In a detailed study of the IllustrisTNG simulations, \citet{Pillepich2018b} found the BCG+ICL fraction in the total cluster stellar content to be around 50 per cent, for objects with mass above $10^{14} M_\odot$. Using the Cluster-EAGLE simulations, \citet{AlonsoAsensio2020} report a fraction $f_\mathrm{ICL} = 0.091 \pm 0.013$ within $R_{200}$. Our result, although in the upper part of the reported range, is still within this range, and thus in overall agreement with previous ICL measurements. 

\subsection{Influence of dynamical state} \label{sec:frac_ds}

In \Fig{fig:massfraction} we studied the ICL mass fraction as a function of cluster mass, and found no significant correlation within the given cluster mass range. We will now investigate the effect of the dynamical state of the clusters. The dynamical state of clusters is a property of special interest, since it can directly affect the cluster mass estimation, and is linked with other properties like halo concentration \citep{Neto2007}. Previous works with \textsc{The Three Hundred} simulations have been devoted to studying the connection between morphology and dynamical state \citep{DeLuca2021,Capalbo2021} or its influence in the hydrostatic mass bias \citep{Gianfagna2023}. The dynamical state is also strongly correlated with the mass accretion history of clusters \citep{Mostoghiu2019,Haggar2020}, and hence its connection with the ICL can provide interesting information regarding the formation channels of the ICL and the cluster itself. Using different sets of simulated galaxy clusters, works like \citet{Rudick2011}, \citet{Cui2014} and \citet{Contini2023a} found more dynamically evolved clusters to have a higher ICL fraction. The opposite interpretation is given in observational studies by \citet{Jimenez-Teja2018,Jimenez-Teja2023}, claiming that higher fractions are expected in merging systems. 
 
There are different ways to quantify the dynamical state of clusters, depending on the nature and purpose of each study. In simulations, \citet{Neto2007} introduced three different parameters to be used as proxies for relaxation of clusters. These parameters are the centre of mass offset, $\Delta_r$, which is the offset between the positions of the centre of mass of the cluster and the density peak, normalised to the halo radius; the subhalo mass fraction, $f_s$, which is the fraction of cluster mass contained in subhalos; and the virial ratio $\eta$, defined as $\eta=(2T-E_s)/ \vert W \vert$, where $T$ is the total kinetic energy of the cluster, $E_s$ its energy from surface pressure and $W$ its total potential energy \citep[see][for an updated calculation for hydrodynamic simulations]{Cui2017}. \citet{Cui2018} used these parameters to study the relaxation of the clusters in \textsc{The Three Hundred}. In a later work, \citet{DeLuca2021} thoroughly investigated the dynamical state of these same clusters, comparing theoretical indicators ($\Delta_r$ and $f_s$) with results from morphology classifications. Also with \textsc{The Three Hundred} clusters, \citet{Haggar2020} introduced the relaxation parameter $\chi_\mathrm{DS}$, that combines the three indicated parameters into a single measure of dynamical state, and used it to correlate with the fraction of backsplash galaxies.

\begin{figure*}
  \centering
  \includegraphics[width=17.7cm]{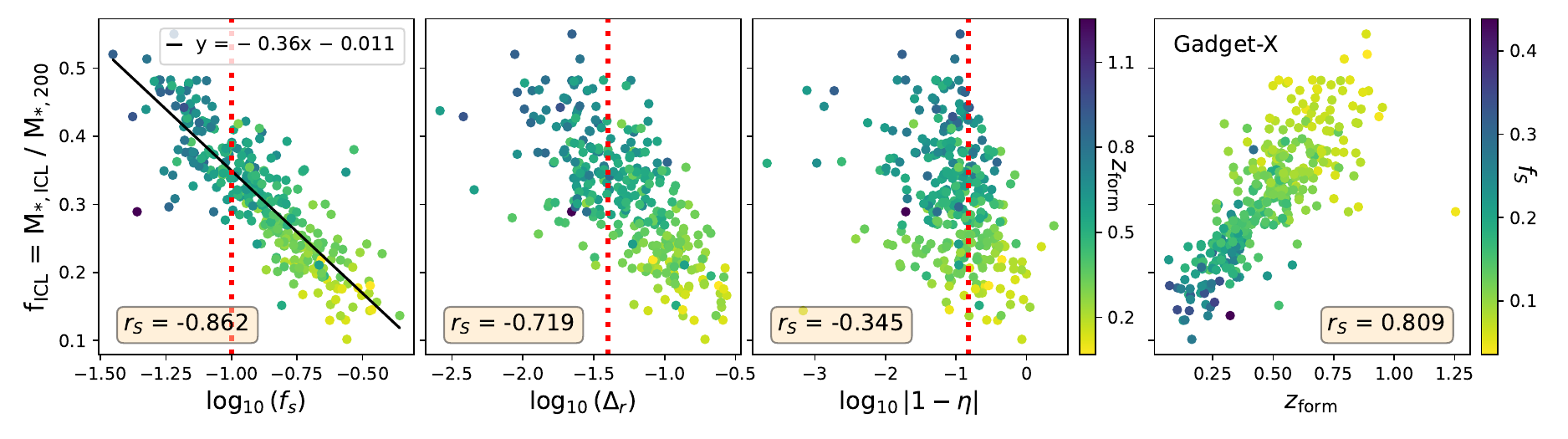}
  \caption{ICL mass fraction within $R_{200}$ of the cluster as a function of the three different theoretical parameters introduced by \citet{Neto2007} to quantify the dynamical state (from left to right subhalo mass fraction $f_s$, centre of mass offset $\Delta_r$ and virial ratio $\eta$), together with the formation time, $z_\mathrm{form}$. The first three panels are coloured by $z_\mathrm{form}$, while the fourth one is coloured by $f_s$, as indicated in the colourbars. In all panels, the Spearman correlation coefficient, $r_S$, is indicated in a lower corner. The vertical dashed lines indicate the thresholds used in \citet{Cui2018} to distinguish between `relaxed' and `unrelaxed' clusters: 0.1, 0.04 and 0.15 respectively. Clusters to the left of these lines (value lower than the threshold) are relaxed, and those to the right are unrelaxed.}
\label{fig:corr_icl}
\end{figure*}

Focusing individually on these three theoretical indicators, we present in \Fig{fig:corr_icl} the ICL mass fraction of each of our 324 clusters as a function of the indicated parameter. From left to right, the first three panels are for subhalo mass fraction $f_s$, centre of mass offset $\Delta_r$ and virial ratio. The first two parameters have to be minimised for the cluster to be most relaxed, while the virial ratio is 1 for a perfectly relaxed cluster (and hence we plot $|1-\eta|$). The vertical dashed lines indicate the thresholds used in \citet{Cui2018} to distinguish between `relaxed' and `unrelaxed' clusters: 0.1, 0.04 and 0.15, respectively. Clusters  with values below the threshold are considered to be relaxed and clusters to the right of the vertical line are labelled as unrelaxed. The value of the Spearman correlation coefficient for each of the plots is indicated in the lower left corner. 

We can see in \Fig{fig:corr_icl} that there is a very clear negative correlation for the first two parameters, meaning that the ICL mass fraction is lower for less relaxed clusters. The correlation coefficient is the highest for the subhalo mass fraction. This could be expected given that the definition of the ICL itself is based on removing the subhalo masses (although for the ICL we consider only stellar mass while $f_s$ takes also DM into account). If a cluster has more mass in subhalos, it will also have more mass in satellites. So, if some of the satellites have not been subject to stripping because they are ``shielded'' by their subhalos, it is likely that the ICL is lower than in a halo with a lower $f_s$.
Given this strong correlation, we fit the points to a straight line and show the results in the first panel of \Fig{fig:corr_icl}, together with the parameters of the best fit, which shows a slope of $-0.36$ (note that $f_s$ is in logarithmic scale in this plot). For the virial ratio we can see that the correlation in the plot is not as clear and the coefficient decreases to -0.35. The same trend remains nevertheless, with the clusters farther from virial equilibrium having in general lower ICL mass fractions.

We also include in the last panel of \Fig{fig:corr_icl} the correlation with the formation time of the cluster, $z_\mathrm{form}$, which, as defined in \citet{Mostoghiu2019}, is computed as the redshift at which $M_{200}$ of the halo is equal to half its value at $z=0$, that is $M_{200}({z_\mathrm{form}})/M_{200}({z=0}) = 0.5$. In this case we see a very clear positive correlation, with $r_S \sim 0.8$, indicating that the earlier the cluster was formed, the higher its ICL mass fraction. Clusters that have accreted much of their mass more recently seem to be in turn dominated by satellite galaxies, which can be indicative of events such as major mergers. These results are in agreement with \citet{Haggar2020}, where a clear correlation is found between formation time and dynamical state of clusters, and with \citet{Contini2023b}, who find a connection between the ICL and the formation time of the cluster itself. These correlations can be further seen in the colour coding of the points in \Fig{fig:corr_icl}. The first three panels are coloured by $z_\mathrm{form}$, while the fourth one is coloured by $f_s$, with the four plots showing clearly that earlier formed clusters (higher $z_\mathrm{form}$) are in general more relaxed. To avoid repetition, we only show here the plots for \textsc{Gadget-X}, but the same general trends hold for \textsc{Gizmo-Simba}.
Moreover, although for simplicity it is not explicitly shown here, we also find a positive correlation between the halo concentration (computed by assuming an NFW profile, \citealp{Navarro1997}) and the ICL fraction, very similarly to the results by \citet{Contini2023a}. This reinforces our findings, given that clusters formed earlier are more concentrated \citep[e.g.][]{Mostoghiu2019}.

\begin{figure}
  \centering
  \includegraphics[width=6.8cm]{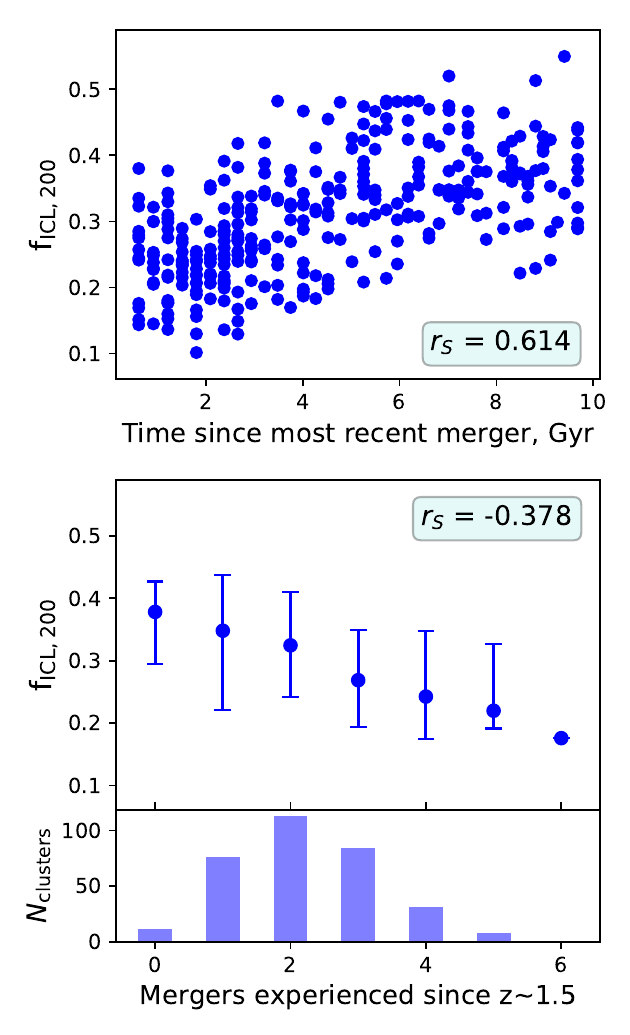}
  \caption{Correlation between ICL mass fraction and merging history of clusters. Top: ICL mass fraction within $R_{200}$ of the cluster as a function of the time since the most recent merger. The mergers are defined following \citet{Contreras-Santos2022a}, as mass increases of at least 25 per cent of the initial mass in less than half a dynamical time. Bottom: ICL mass fraction as a function of the number of mergers experienced by the cluster since $z \sim 1.5$. The clusters are binned by number of mergers experienced, with the bar plot below indicating the number of clusters in each bin. The dots correspond to the median values, while the errors are 16th-84th percentiles.}
\label{fig:corr_icl_mergers}
\end{figure}

To further explore the physical situation behind \Fig{fig:corr_icl} and quantify the dynamical state of clusters in a different way, we now use the information from a previous work about galaxy cluster mergers in \textsc{The Three Hundred} simulations \citep{Contreras-Santos2022a}. In this work, we defined mergers as significant mass increases happening in less than half a dynamical time of the cluster. We apply now this same procedure and consider mergers as increases of at least 25 per cent in the total mass of the cluster\footnote{Although we do not explicitly compute the ratio of the two merging objects \citep{Contreras-Santos2022a}, these mergers can be considered as `major' in the sense that they result in a significant increase in the mass of the cluster (by at least 25 per cent).}, that is $(M_f-M_i)/M_i \geq 0.25$, where $M_f$ and $M_i$ are the masses at two different snapshots separated by half a dynamical time. 
In the top panel of \Fig{fig:corr_icl_mergers} we plot the ICL fraction against the time since the last merger undergone by the cluster. Although the scatter is high, we find a positive correlation with Spearman coefficient $r_S \sim 0.6$, so that clusters that have recently undergone a merger have a lower ICL fraction, in agreement with our previous findings. The lower panel in \Fig{fig:corr_icl_mergers} shows also the ICL fraction, but now as a function of the number of mergers undergone by the cluster since $z\sim 1.5$. The clusters are binned for each value between 0 and 6 mergers, the dots showing the median value and the errors computed as 16th-84th percentiles. The bar plot in the bottom shows the number of clusters within each bin. We see here that, although the scatter is still high and the correlation is weaker, there is also a trend for clusters that have recently experienced more mergers to have a smaller ICL fraction. 

Overall, our results point to a scenario where more relaxed clusters have a higher ICL fraction than disturbed or merging clusters, that have formed more recently. \citet[][see also \citealp{Contini2023b} for the extension of these results to $z=2$]{Contini2023a} propose a very similar scenario, where the main driver of ICL formation is the concentration, that also separates dynamically evolved clusters (more concentrated) from younger clusters (less concentrated). Our results are also in agreement with observational findings by \citet[][see also \citealp{DaRocha2008,Poliakov2021,Ragusa2022}]{Ragusa2023}, who find that clusters with a higher fraction of early type galaxies (more relaxed) show higher ICL fractions. 
Nevertheless, recent results by \citet[][see also \citealp{Jimenez-Teja2018,deOliveira2022}]{Jimenez-Teja2023} support the opposite situation, where merging clusters present a higher ICL fraction than more passive and relaxed ones. This highlights that the questions of the origin and formation of the ICL and its connection to the formation of the cluster are still open and remain to be understood. For this paper we prefer to keep our study to $z=0$, while we will go into much detail about where the ICL stars are coming from in a future paper already in preparation.

\subsection{Selecting the most relaxed and unrelaxed clusters} \label{sec:frac_ds_sep}
We have seen that, while the cluster mass has no influence on the ICL mass fraction, this fraction shows a clear correlation with the dynamical state of the cluster, such that more relaxed clusters have a higher ICL fraction. We now want to further quantify this effect. For this we will select the most relaxed and most unrelaxed clusters in the sample, and study how their stellar mass is distributed among the different components. Following \citet{Haggar2020}, in order to obtain a continuous, non-binary measure of dynamical state, we combine the three theoretical parameters previously mentioned into a single measure of dynamical state, the so-called `relaxation parameter' of a cluster:
\begin{equation}
\chi_{\mathrm{DS}}=\sqrt{\frac{3}{\left(\frac{\Delta_r}{0.04}\right)^2+\left(\frac{f_s}{0.1}\right)^2+\left(\frac{\vert 1 - \eta \vert}{0.15}\right)^2}}.
\label{eq:chiDS}
\end{equation}
This parameter satisfies that, for `dynamically relaxed' clusters, $\chi_{\mathrm{DS}} \gtrsim 1$, and the higher $\chi_{\mathrm{DS}}$ the more relaxed a cluster is.

\begin{figure}
  \centering
  \includegraphics[width=7cm]{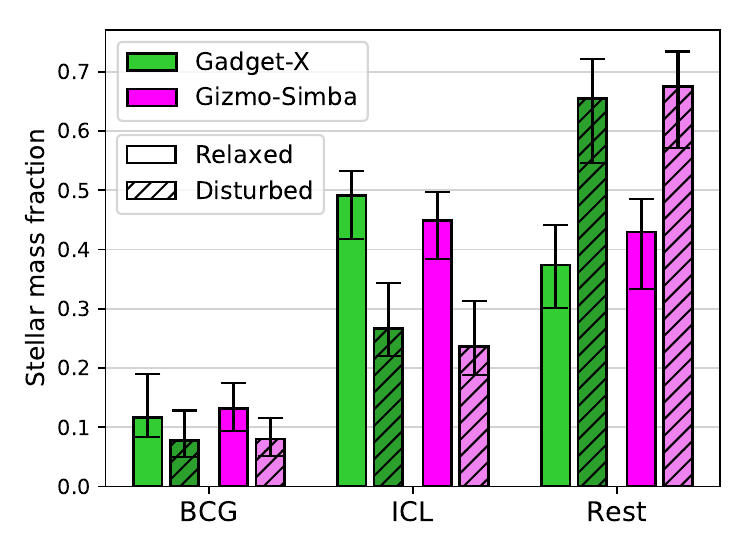}
  \caption{Stellar mass fraction for the BCG, ICL and rest of stars. The bar shows the median value for the 324 clusters, while the error bars indicate 16th-84th percentiles. The fraction is computed dividing the stellar mass content in each of the components by the total stellar mass inside $R_{500}$, $M_{*,500}$. The straight and dashed bars separate into relaxed and disturbed clusters based on their $\chi_\mathrm{DS}$ value at $R_{500}$. `Relaxed' clusters are the 50 clusters with the highest $\chi_\mathrm{DS}$, while `disturbed' are the 50 clusters with the lowest value of $\chi_\mathrm{DS}$. As before, green bars are for \textsc{Gadget-X}, while magenta ones are for \textsc{Gizmo-Simba} results.}
\label{fig:fractions_barplot_sep}
\end{figure}

Using this parameter we can easily select the most relaxed clusters within our sample, together with the most unrelaxed or disturbed. Given the size of \textsc{The Three Hundred} sample, that allows us to make subsamples that are still statistically significant, we will select the 50 most extreme clusters in both sides. In \Fig{fig:fractions_barplot_sep} we show, similarly to \Fig{fig:fractions_barplot}, the stellar mass fraction in the different components: BCG, ICL and the rest of the galaxies in the clusters. But in this plot it is shown only for the subsamples of the 50 most relaxed and disturbed clusters, that is those with, respectively, highest and lowest $\chi_\mathrm{DS}$. The bars indicate the median values, while the error bars show the 16th-84th percentiles. The different colours designate the different codes, green for \textsc{Gadget-X} and magenta for \textsc{Gizmo-Simba}. \Fig{fig:fractions_barplot_sep} shows very clearly the effect we have previously discussed, with relaxed clusters having a higher ICL fraction. It also highlights how the stellar mass content in satellite galaxies is significantly higher for disturbed clusters. For the BCGs, the error bars show that the difference between relaxed and disturbed clusters is less significant, but we still see a change from $\sim12\%$ in relaxed clusters to $\sim 7\%$ for disturbed ones (almost 50 per cent difference between them). From a different perspective, \Fig{fig:fractions_barplot_sep} shows how the ICL can be used as an indicator of the cluster dynamical state: relaxed clusters can be identified as those with $f_\mathrm{ICL} > f\mathrm{rest}$.


In conclusion, and focusing on the ICL, we find that for our whole cluster sample the ICL mass fraction, considering 16th-84th percentiles as the errors, is $0.38 \pm 0.10$ ($0.34 \pm 0.09$) for \textsc{Gadget-X} (\textsc{Gizmo-Simba}). When selecting a subsample of the 50 most relaxed clusters, this fraction increases to $0.49_{-0.06}^{+0.05}$ ($0.45_{-0.07}^{+0.05}$); while it decreases to $0.27_{-0.06}^{+0.08}$ ($0.24_{-0.04}^{+0.08}$) for the 50 most disturbed clusters, according to their $\chi_\mathrm{DS}$ value at $z=0$. We note that these fractions are in perfect agreement with those found by \citet{Contini2023a} for the most and least concentrated halos in their sample. We explain these differences between the subsamples by their different formation scenarios, with more disturbed clusters being the result of recent merger events, such that galaxies as a whole enter the cluster and remain as bound objects; and relaxed clusters being the result of the disruption of these galaxies and hence slow accretion of their stars into the ICL.
In agreement with our results, but with different simulations, \citet{Cui2014} and \citet{Rudick2011} found the ICL fraction to be higher for relaxed clusters, and stated that this can be understood if relaxed clusters are more dynamically evolved than disturbed ones, with the diffuse star particles mainly coming from satellite galaxies undergoing mergers and stripping. This can also be understood within the context of the two-phase formation scenario of galaxy clusters (e.g. \citealp{GunnGott1972,Gunn1977,Ascasibar2004}), that describes an early fast accretion phase building up the central region; followed by a slow accretion phase, where the inner part remains more constant and the mass builds up in the outer regions. This was analysed by \citet{Mostoghiu2019} in \textsc{The Three Hundred} clusters, concluding that unrelaxed clusters are still in their fast accretion mode, while relaxed clusters have already reached the slower phase.
We plan to investigate the origin of the ICL in more detail in a future work, where we will trace the particles back in time. Nevertheless, for the present work we prefer to focus on characterising the ICL at present day, and how it is related with other cluster properties. 

\section{Relation of ICL to dark matter} \label{sec:icl_dm}

In this section we study the relationship between the ICL and the dark matter (DM) component of galaxy clusters. This relationship is particularly interesting since it could provide a way of exploring the DM component by using the ICL, which means probing the DM in clusters using only deep imaging observations. This has been previously investigated using Hubble Space Telescope (HST) data by \citet{MontesTrujillo2019}, finding very promising results pointing to a very similar matter distribution of both components. From the simulations side there has also been a great effort to separate the stellar and DM component of halos, in order to understand their origins and joint evolution. \citet{Libeskind2011} used constrained simulations of the local Universe, while \citet{Pillepich2018b} worked with the cosmological simulation IllustrisTNG. Working with the Cluster-EAGLE simulations, \citet{AlonsoAsensio2020} and \citet{Deason2021} also found that the ICL can be used to infer properties of the underlying DM halo, like its shape and boundaries.

\subsection{Full cluster sample}

\begin{figure*}
  \centering
  \includegraphics[width=14cm]{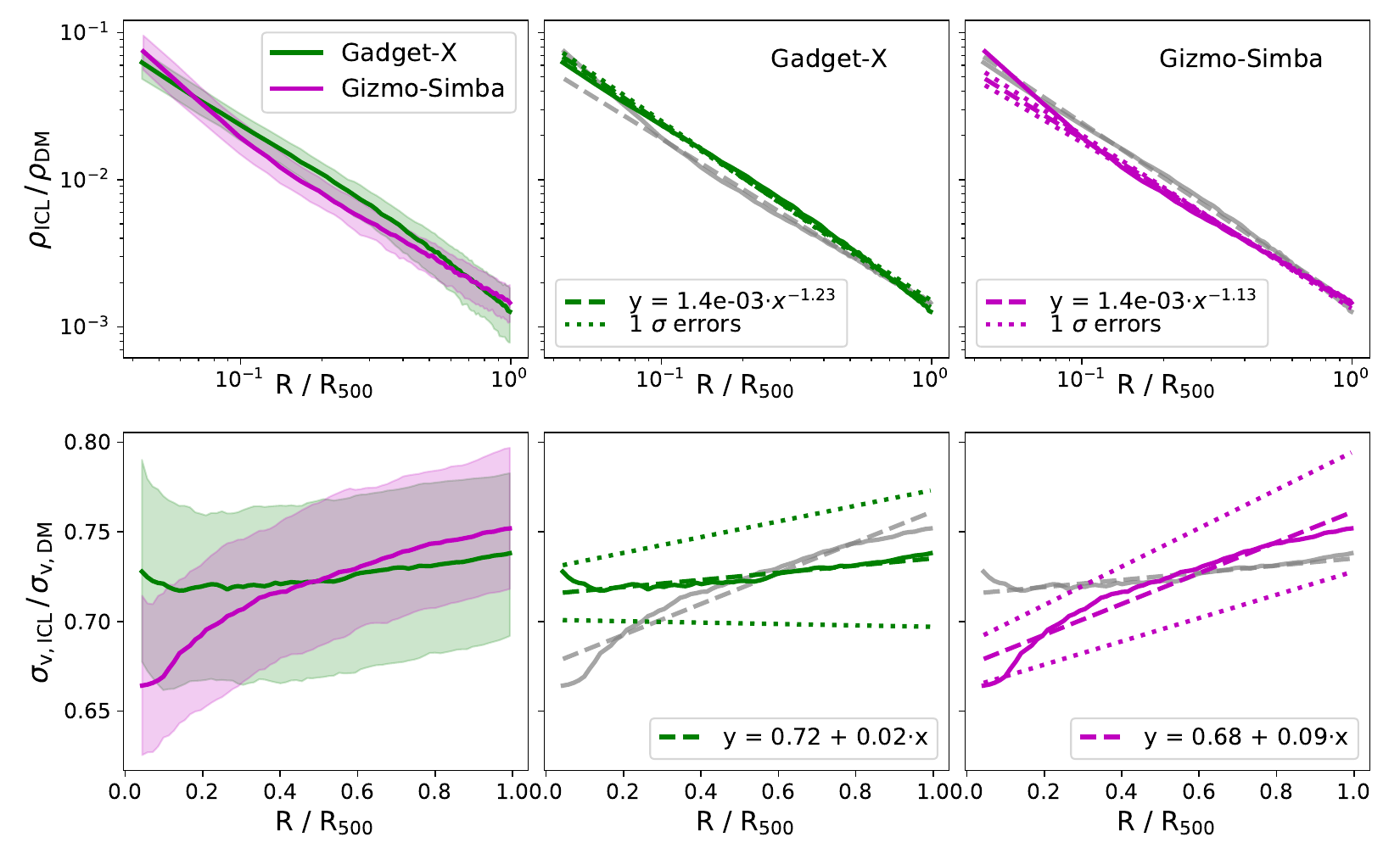}
  \caption{Relation between ICL and DM radial profiles. Top: ratio between the ICL and DM density profiles. In the left panel, solid lines are the median values, green for \textsc{Gadget-X} and magenta for \textsc{Gizmo-Simba}, while the shaded regions are the 16th-84th percentiles. The middle and right panels show the best fit to an exponential curve of the \textsc{Gadget-X} and \textsc{Gizmo-Simba} ratios, respectively (the fainter grey lines are for the opposite simulation, for easier comparison between them). The dashed lines depict the best fit (the analytical formula can be seen in the legend), while the dotted lines indicate the 1 $\sigma$ errors. Bottom: same as the top row but for the ratio between ICL and DM velocity dispersion profiles.}
\label{fig:icldm_fits}
\end{figure*}

In order to explore this relationship, we have computed the volumetric density profile and the velocity dispersion profiles for both the DM and ICL components. The density profile for the ICL is very similar to that in \Fig{fig:bcgicl_profile}, with the difference that now we are excluding the BCG. For the DM we compute this profile by considering all the DM particles that belong to the central halo. In \Fig{fig:icldm_fits}, the top left panel shows the ratio between these two profiles, $\rho_\mathrm{ICL}/\rho_\mathrm{DM}$, stacked for all the 324 clusters. The solid lines are obtained by computing the ratio $\rho_\mathrm{ICL}/\rho_\mathrm{DM}$ at each $R$ for all the 324 clusters and then computing the median value. The shaded regions show the 16th-84th percentiles, green is for \textsc{Gadget-X} and magenta for \textsc{Gizmo-Simba}. We can see that the results are very similar for both simulations, with the ratio showing a reasonably constant slope across different radii, when using logarithmic scale in both axes. The plot in the middle of the upper row shows how this ratio $\rho_\mathrm{ICL}/\rho_\mathrm{DM}$ can be fitted to a power law of the form $y=a \cdot x^b$. The dashed line shows the best fit, while the dotted lines show the 1 $\sigma$ errors. The green lines correspond to the results for \textsc{Gadget-X}, while the ratio and best fit curve for \textsc{Gizmo-Simba} are shown in grey, to facilitate the comparison between the two simulations. The plot in the upper right corner is the same as the central one but for \textsc{Gizmo-Simba}, with the results for \textsc{Gadget-X} in grey. The legends in the lower part of the plots indicate the parameters of the fit, which are also summarised in \Tab{table:fit_params}. We want to highlight here the value of the exponent $b$, which determines the slope of the curves in the upper row of \Fig{fig:icldm_fits}. This parameter takes a value of $-1.23 \pm 0.04$ ($-1.13 \pm 0.05$) for \textsc{Gadget-X} (\textsc{Gizmo-Simba}). We note that these results have been obtained considering all the DM particles, that is including also the substructures present in the halo. We have compared the results for both possibilities (excluding or including the substructures) and the conclusions are very similar, with the parameters being compatible within 1 $\sigma$ with each other. For the remaining of this section, the DM mass will refer to the inclusive one, considering all the mass in subhalos as well.

Also regarding the values in \Tab{table:fit_params}, we want to clarify that the parameters --as well as the errors-- have been obtained by stacking all the $\rho_\mathrm{ICL}/\rho_\mathrm{DM}$ profiles for the 324 clusters and then fitting to this one profile. An alternative way of obtaining these values is to fit the profile of each individual cluster and obtain the median best-fit parameters with their percentiles. We prefer the former option since it provides a more robust fit that focuses on the overall trend for our sample, and hence reveals general properties of the relation between ICL and DM. The latter option, that is, fitting for each cluster, considers more the individual variations among clusters and provides cluster-specific values. Although not shown here, we have applied this methodology, and find that the median values of the parameters remain within the 1 $\sigma$ regions, but the errors are considerably larger (3-4 times larger than for the stacked profile for parameter $b$ and 2-3 times for parameter $d$).


\begin{table}
\centering
\caption{Parameters of the best fit curves to the ratios between the ICL and DM density and velocity dispersion profiles shown in \Fig{fig:icldm_fits}, together with their 1 $\sigma$ errors.}
\begin{tabular}{r c c c}
\hline
    &  &  {\textbf{\textsc{Gadget-X}}} & {\textbf{\textsc{Gizmo-Simba}}} \\ \hline
  \bm{$\rho_\mathrm{ICL}$} \textbf{/} \bm{$\rho_\mathrm{DM}$}  & $\bm{a}(/10^{-3})$  & $1.42 \pm 0.08$ &  $1.39 \pm 0.07$    \\ 
                         $y = a \cdot x^b$          & $\bm{b}$  & $-1.23 \pm 0.04$    &  $-1.13 \pm 0.05$       \\ \hline
  \bm{$\sigma_\mathrm{v,ICL}$} \textbf{/} \bm{$\sigma_\mathrm{v,DM}$}  & $\bm{c}$  & $0.72 \pm 0.01$ &  $0.68 \pm 0.01$    \\ 
                         $y = c + d \cdot x$          & $\bm{d}$  & $0.02 \pm 0.02$    &  $0.09 \pm 0.02$ \\ \hline
\end{tabular}
\label{table:fit_params}
\end{table}

A study by \citet{Deason2021} also explores the connection between ICL and DM in the Cluster-EAGLE simulations but focusing on the `edge' of their distributions, finding that they are both closely related. In that work (cf. their Fig. 4) it can be seen that the logarithmic slope profiles for DM and stars, when stacking all the clusters, show an approximately constant difference between them, which is the same we see in our \Fig{fig:icldm_fits}. Computing the difference between the two logarithmic profiles, the value in \citet{Deason2021} is $\sim -1$, in agreement with the value for the slope $b$ we find (see \Tab{table:fit_params}).

\begin{figure*}
  \centering
  \vspace{0.3cm}
  \includegraphics[width=16.5cm]{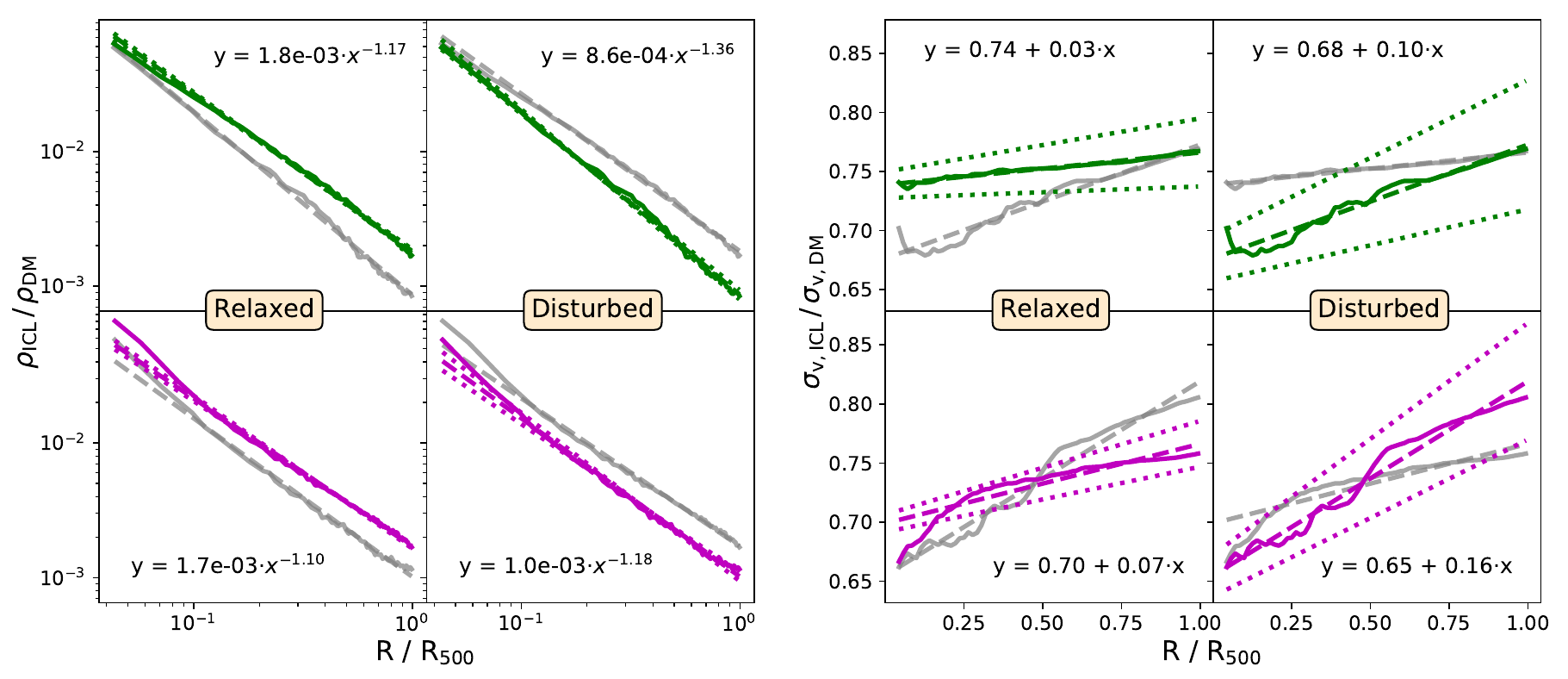}
  \caption{For the density (left) and velocity dispersion (right) profiles, ratio between ICL and DM components. Same as \Fig{fig:icldm_fits} but for the subsamples of the 50 most relaxed and disturbed clusters (see \Sec{sec:frac_ds_sep}). Green is for \textsc{Gadget-X} and magenta for \textsc{Gizmo-Simba} simulations. In each panel, the grey line corresponds to the opposite subsample for the same simulation, namely relaxed or disturbed, to allow for an easier comparison between relaxed and disturbed clusters.}
\label{fig:icldm_fits_sep}
\end{figure*}

The bottom row of \Fig{fig:icldm_fits} focuses on the velocity dispersion profiles of both the DM and ICL components. We compute this profile as the velocity dispersion for all the particles inside each indicated radius (instead of the particles within a thin spherical shell). As for the density profile, in the left panel we show the ratio between the profiles for ICL and DM, $\sigma_\mathrm{v,ICL}/\sigma_\mathrm{v,DM}$. We see again that the two simulations agree across the majority of radii, and show only slight differences in the innermost part of the cluster, a feature we have already seen and explained in this paper (e.g. \Fig{fig:bcgicl_profile}). In the two other panels we can see how this ratio can be fitted to a straight line with the form $y = c+d \cdot x$ (note that the scales are not logarithmic in this case). Again, the values of these free parameters are summarised in \Tab{table:fit_params}. The value obtained for the slope is $0.02 \pm 0.02$ ($0.09 \pm 0.02$) for \textsc{Gadget-X} (\textsc{Gizmo-Simba}). Both values are significantly low, and even compatible with 0 for \textsc{Gadget-X}, meaning that the ICL and the DM components have velocity dispersion profiles that evolve very similarly.

In a previous study, \citet{MontesTrujillo2019} used deep observations of galaxy clusters to show that the projected ICL distribution closely follows that of the DM distribution up to $\sim 140$ kpc, thus providing an accurate luminous tracer of DM. This was confirmed by \citet{AlonsoAsensio2020} in the Cluster-EAGLE simulations, where they also found that the ratio between the ICL and the total mass density profile has a slope of $\sim -1$, in agreement with what we find for \textsc{The Three Hundred} simulations (\Fig{fig:icldm_fits}). We still want to note that the two studies mentioned work with projected rather than volumetric quantities, and so the comparison with them should be interpreted with caution. With a different set of simulations, \citet{Yoo2022} developed a new method to measure the similarity between two distributions, further supporting these findings. This result regarding the relation between ICL and DM means that the ICL density profile can be used to infer the DM density profile, which can in turn be integrated to obtain an estimation of the cluster mass. \citet{Pillepich2018b} propose a similar method of inferring halo mass from stellar mass based on the IllustrisTNG simulations. 


We further explore the relationship between these two components by introducing the ratio between the ICL and DM velocity dispersion profiles. The velocity dispersion is a very interesting property of clusters, since it can be used as a proxy of cluster mass (see e.g. \citealp{Allen2011}), and thus for inferring cosmological parameters. Several studies have been devoted to studying this relation between velocity dispersion and cluster mass, and the differences between using the member galaxies or the DM component (see \citealp{Munari2013} or \citealp{Ferragamo2022} for a study in \textsc{The Three Hundred} clusters).
In turn, the velocity dispersion of the ICL stars can be a useful tool to understand in more detail the kinematics of the cluster, and thus its structure and evolution. Some recent works already present measurements for nearby clusters in the Local Universe (see e.g. \citealp{Spiniello2018} for the Fornax cluster or \citealp{Arnaboldi2022} for a review). In this work we present the relation between the velocity dispersion of the DM and that of the diffuse stars that follow the gravitational potential of the cluster. We find that the ratio ICL/DM can be described by a straight line with slope close to 0, meaning that they evolve very similarly up to $R_{500}$, so that also one of them can be used to infer the other one (see equation and parameters in \Tab{table:fit_params}).

\subsection{Subsamples of relaxed and disturbed clusters}

In the previous \Sec{sec:frac_ds_sep} we created two subsamples of the 50 most relaxed and disturbed clusters based on their value of the relaxation parameter $\chi_\mathrm{DS}$, and we studied how their compositions differed from each other. We found that the ICL fraction is significantly higher for the relaxed clusters, for which this component can even dominate the stellar component of clusters. Now that we have studied the relation of the ICL with the DM component, we also ask ourselves the question of whether this relation depends on the dynamical state of the cluster. In order to check this, we have repeated the previous analysis, this is, the comparison between density and velocity dispersion profiles for ICL and DM components, for these two subsamples of relaxed and disturbed clusters.

We show the results in \Fig{fig:icldm_fits_sep}. The left $2 \times 2$ panel in this figure is for the density, while the right one is for the velocity dispersion profiles. In each of these $2 \times 2$ panels, the first column is for the relaxed clusters, \textsc{Gadget-X} on top (in green) and \textsc{Gizmo-Simba} on the bottom (in magenta); while the second column is for the disturbed clusters. Similarly to the second and third columns of \Fig{fig:icldm_fits}, in \Fig{fig:icldm_fits_sep} we show the median value (solid line) of the ratio of the two profiles, $\rho_\mathrm{ICL}/\rho_\mathrm{DM}$ or $\sigma_\mathrm{v,ICL}/\sigma_\mathrm{v,DM}$, together with the best fit curve (dashed line) and the 1 $\sigma$ regions for the fits (delimited by dotted lines). In each panel, the grey line corresponds to the opposite subsample for the same simulation, for example in the upper left corner the green lines correspond to `relaxed \textsc{Gadget-X}' clusters, and so the grey lines are the median and best fit for the `disturbed \textsc{Gadget-X}' clusters. This allows for an easier comparison between the relaxed and disturbed subsamples, which is the main goal of this subsection. The equations that describe the fits are indicated in the plot, while the values of the parameters with their 1 $\sigma$ errors are summarised in \Tab{table:fit_params_sep}.

Focusing first on the density, we see in \Fig{fig:icldm_fits_sep} that for relaxed clusters the line is moved up with respect to the disturbed ones, which is a consequence of the ICL fraction being higher for relaxed clusters (see \Sec{sec:icl_frac}). Looking at the values of the exponents of the fits, parameter $b$ in \Tab{table:fit_params_sep}, we see that the ratio becomes a bit steeper for the disturbed clusters, although the error is also higher for these clusters and hence the difference is not very significant. Regarding the velocity dispersion, \Fig{fig:icldm_fits_sep} shows a more relevant difference between relaxed and disturbed clusters. For relaxed clusters the ratio remains almost constant, while the slope is significantly increased for disturbed ones. This difference is more clear for \textsc{Gadget-X}, but is present in both simulations. Parameter $d$ in \Tab{table:fit_params_sep} quantifies the slope of this ratio, that for relaxed clusters is compatible with 0 for \textsc{Gadget-X} and a bit higher for \textsc{Gizmo-Simba}, but still below the value for the whole cluster sample. This can be explained by the fact that, for relaxed clusters, there are less substructures, less mergers happening and the mass is more concentrated towards the centre. These factors make the velocity dispersion to remain more or less constant beyond a certain point in the profile. For disturbed clusters, substructure and ongoing mergers can cause the velocity dispersion to continue growing with increasing distance from the cluster centre. This effect is more pronounced for the ICL than for the DM, which is less affected by interactions, and hence we see an effect in the ratio $\sigma_\mathrm{v,ICL}/\sigma_\mathrm{v,DM}$. 
In the context of our previous conclusions, that relaxed clusters are more dynamically evolved than disturbed ones, this $\sigma_\mathrm{v}$ ratio is constant for relaxed clusters because accreted stars have already had time to relax and trace the underlying gravitational potential. However, for unrelaxed clusters, if the ICL is formed of ejected stars, these have just been stripped from satellites and are not yet fully coupled and in equilibrium with the halo potential.

We have seen in this subsection how the relation between ICL and DM is also affected by the dynamical state of the cluster, affecting both the density and the velocity dispersion profiles, but more importantly the latter. Since the velocity dispersion profile of the ICL can be influenced by many factors, the exact relationship with the dynamical state can be complex and different from cluster to cluster. We have presented here our median results for subsamples of 50 of the most relaxed and disturbed clusters.


\begin{table}
\centering
\caption{Same as \Tab{table:fit_params} but for the subsamples of the most relaxed and disturbed clusters. Parameters of the best fit curves to the ratios between the ICL and DM density and velocity dispersion profiles shown in \Fig{fig:icldm_fits_sep}, together with their 1 $\sigma$ errors.}
\begin{tabular}{rl c c}
\hline
    &  &  {\textbf{Relaxed}} & {\textbf{Disturbed}} \\ \hline
    $\bm{a}(/10^{-3})$ & \textsc{Gadget-X}       & $1.81 \pm 0.06$ &  $0.86 \pm 0.07$    \\ 
          & \textsc{Gizmo-Simba}    & $1.70 \pm 0.06$ &  $1.01 \pm 0.07$    \\ 
    $\bm{b}$ & \textsc{Gadget-X}       & $-1.17 \pm 0.03$    &  $-1.36 \pm 0.05$       \\ 
             & \textsc{Gizmo-Simba}    & $-1.10 \pm 0.04$    &  $-1.18 \pm 0.07$ \\ \hline
    $\bm{c}$ & \textsc{Gadget-X}       & $0.74 \pm 0.01$    &  $0.68 \pm 0.02$    \\ 
             & \textsc{Gizmo-Simba}    & $0.70 \pm 0.01$    &  $0.65 \pm 0.02$    \\ 
    $\bm{d}$ & \textsc{Gadget-X}       & $0.03 \pm 0.02$    &  $0.10 \pm 0.04$       \\ 
             & \textsc{Gizmo-Simba}    & $0.07 \pm 0.01$    &  $0.16 \pm 0.03$ \\ \hline
\end{tabular}
\tablefoot{For the density we fit to a power law $y = a \cdot x^b$, while for the velocity dispersion we use a straight line $y = c + d \cdot x$.}
\label{table:fit_params_sep}
\end{table}

\section{Conclusions} \label{sec:conclusions}

In this work, we present the first detailed analysis of the intra-cluster light (ICL) within \textsc{The Three Hundred} simulations project. This consists of a suite of 324 numerically modelled spherical regions centred on the most massive clusters found in a prior DM-only cosmological simulation. As such, it makes a relatively large set of galaxy clusters compared to other state-of-the-art cluster simulations, which typically focus on a handful of objects (for instance, the Cluster-EAGLE simulations, \citealp{Barnes2017}, include only 30 galaxy clusters). \textsc{The Three Hundred} clusters constitute a mass-complete sample within the range $14.4< \log(M_{500}/M_\odot)<15.4$ at $z=0$. The size of the sample allows us to perform a statistical study of the ICL in clusters, as well as to make statistically significant subsamples focusing on different cluster properties.
These 324 regions of 15 $\hMpc$ in radius around the cluster centre have been resimulated including full hydrodynamics with two different subgrid physics implementations: $\textsc{Gadget-X}$, which uses a modified version of the non-public \textsc{Gadget3} code \citep{Murante2007,Rasia2015}; and \textsc{Gizmo-Simba}, performed with the \textsc{Gizmo} code \citep{Hopkins2015} and the galaxy formation subgrid models from the \textsc{Simba} simulation \citep{Dave2019}. Comparing the results for the two simulations can help to understand the effects of different physical processes and test the robustness of the resulting predictions.

We first define the BCG as the total stellar mass contained within a fixed spherical aperture of 50 kpc (in \Sec{sec:id_aperturesize} and \App{appendix:aperture} we study the dependence on the size of this aperture) from the halo centre (located by the AHF halo finder as a peak in the density field). This is a common approach to find the BCG in numerical simulations, with both theoretical and observational arguments supporting it \citep{Pillepich2018b,Kravtsov2018}. Then, from the rest of the stellar particles, we find the ICL by selecting those that do not belong to any subhalo. In this way we remove the central and the satellite galaxies, leaving us with the particles that are bound to the potential of the cluster itself, but neither the BCG nor any of the satellite galaxies. This diffuse stellar component is generally referred to as ICL (see \Fig{fig:icl_example} for an example representation for one cluster). 

As a first check, in \Sec{sec:bcg_icl} we compare our resulting BCG+ICL mass to observational results by \citet{Gonzalez2013}, \citet{Kravtsov2018} and \citet{DeMaio2020} and find our predictions to be in overall agreement with them. The two simulations \textsc{Gadget-X} and \textsc{Gizmo-Simba} show highly consistent results down to 0.01 $R_{500}$. Inside this region, the density profile from \textsc{Gadget-X} shows a steeper slope, while \textsc{Gizmo-Simba} starts to flatten. Although these scales are very close to the simulations resolution limit, the differences between the two simulations can be attributed to the different stellar formation and AGN implementation (cf. \Sec{sec:data}), which eventually leads to higher stellar masses in the very central regions found in the \textsc{Gadget-X} runs.

In Sections \ref{sec:icl_frac} and~\ref{sec:icl_dm} we solely focus on the ICL of \textsc{The Three Hundred} clusters. We study its mass fraction and how it depends on halo mass and dynamical state. We also study the relationship between the ICL and the DM components of clusters, verifying whether or not the former could be used to trace the latter. The main results of these sections, which are the main findings of this work, are summarised as follows.

\begin{itemize}
    \item The ICL mass fraction, computed as $M_\mathrm{ICL,\Delta}/M_{*,\Delta}$ (with $\Delta = 200, 500$), has a median value of $f_\mathrm{ICL,500} = 0.36$ and $f_\mathrm{ICL,200} = 0.29$, with 1 $\sigma$ error of $\pm 0.10$. This value is independent of halo mass, at least in our considered mass range (\Fig{fig:massfraction}).
    \item The ICL mass fraction shows a clear dependence on the dynamical state of the cluster, which is quantified by the theoretical indicators introduced by \citet{Neto2007}: subhalo mass fraction $f_s$, centre of mass offset $\Delta_r$ and -- to a lesser extent -- with the virial ratio $\eta$. With a different approach and based on our previous work \citep{Contreras-Santos2022a}, we find a correlation between the ICL fraction and the merging history of the cluster. Clusters that have undergone their last merger more recently present a smaller fraction. There is also a weaker negative correlation between the number of mergers undergone since $z \sim 1.5$ and the ICL fraction (see \Fig{fig:corr_icl_mergers}). These results are in agreement with the scenario where relaxed clusters are dynamically more evolved than disturbed ones, and hence star particles in the ICL component mainly come from the satellite galaxies undergoing merging and tidal stripping \citep{Rudick2011,Cui2014,Iodice2020,Ragusa2021,Ragusa2023,Contini2023a}.
    \item The ratio between the volumetric density profiles of the ICL and the DM components follows a power law up to $R_{500}$, with exponent $-1.23 \pm 0.04$ ($-1.13 \pm 0.05$) for \textsc{Gadget-X} (\textsc{Gizmo-Simba}). The full equations, together with the 1 $\sigma$ uncertainties of each parameter, can be seen in \Tab{table:fit_params}. As suggested in previous works \citep{Pillepich2018b,AlonsoAsensio2020}, this relation between the density profiles can be used as a method to infer the DM mass of the halo. Our study, with more than three hundred clusters, provides a robust relation and confirmation of this method.
    \item The ratio between the velocity dispersion profiles of the ICL and the DM component follows a straight line with slope close to 0, $0.02 \pm 0.02$ for \textsc{Gadget-X} and $0.09 \pm 0.02$ for \textsc{Gizmo-Simba} (see \Tab{table:fit_params} for the full equations). This relation can also be used to infer the DM halo velocity dispersion up to $R_{500}$, which evolves very similarly to that of the ICL. 
\end{itemize}

One limitation of this study that needs to be mentioned is in regard to the selection of the cluster centre, which is in turn used to find the BCG. The halo finder applied by us, namely AHF, locates the halo centre as a peak in the density field, so that we always centre our cluster in the highest density peak, which is in general the BCG. However, some studies have shown that BCGs are not always located at the centre of their clusters, especially when the cluster is unrelaxed \citep{Martel2014,DePropris2021}. If the BCG is not in the centre, some of its particles might be mislabelled as ICL, and vice versa. Nevertheless, we do not expect these particles to be a majority, especially for the ICL, which is considerably more massive than the BCG, and so we believe our results for the ICL to still be consistent in spite of this.


Finally, we are already planning a future work to study the origin of the ICL in more detail by tracing the particles back in time and investigating where they were formed. Similarly to the work with the IllustrisTNG simulation by \citet{Montenegro-Taborda2023}, we will separate the stars into \textit{in situ} and \textit{ex situ}, depending on whether they were formed in the ICL itself or somewhere else and then accreted by this diffuse component. This will help us to better understand the relation between the DM component and the ICL in clusters, and in general to gain a better understanding of the process of galaxy cluster formation and evolution.

\begin{acknowledgements}
    The authors would like to thank Emanuele Contini for carefully reading the manuscript and providing insightful suggestions. This work has been made possible by \textsc{The Three Hundred} (\url{https://the300-project.org}) collaboration. The simulations used in this paper have been performed in the MareNostrum Supercomputer at the Barcelona Supercomputing Center, thanks to CPU time granted by the Red Espa\~{n}ola de Supercomputaci\'on. As part of \textsc{The Three Hundred} project, this work has received financial support from the European Union’s Horizon 2020 Research and Innovation programme under the Marie Sklodowskaw-Curie grant agreement number 734374, the LACEGAL project. 
    ACS, AK, IAA and CDV thank the Ministerio de Ciencia e Innovaci\'{o}n (MICINN) for financial support under research grants PID2021-122603NB-C21 and PID2021-122603NB-C22. AK further thanks Drop Nineteens for `kick the tragedy'. WC is supported by the STFC AGP Grant ST/V000594/1 and the Atracci\'{o}n de Talento Contract no. 2020-T1/TIC-19882 granted by the Comunidad de Madrid in Spain. He also thanks the Ministerio de Ciencia e Innovación (Spain) for financial support under Project grant PID2021-122603NB-C21 and ERC: HORIZON-TMA-MSCA-SE for supporting the LACEGAL-III project with grant number 101086388. 
\end{acknowledgements}



\bibliographystyle{aa}
\bibliography{archive}

\appendix

\section{Dependence on aperture size}  \label{appendix:aperture}

In this section we show how the main results shown throughout the paper are affected by the choice of aperture size to separate BCG and ICL. In the previous sections we have used an aperture of 50 kpc from the halo centre to select all the stars that belong to the BCG. Now, we will also use values of 30 and 70 kpc, to see how much the results are influenced by a smaller or larger aperture. Here we like to remind the reader that we apply a 3D aperture, and hence this method is not fully reproducible in observations (see discussion in Sec.~\ref{sec:identification:bcg}).

\subsection{ICL mass fraction}

\begin{figure}
  \centering
  \includegraphics[width=6.7cm]{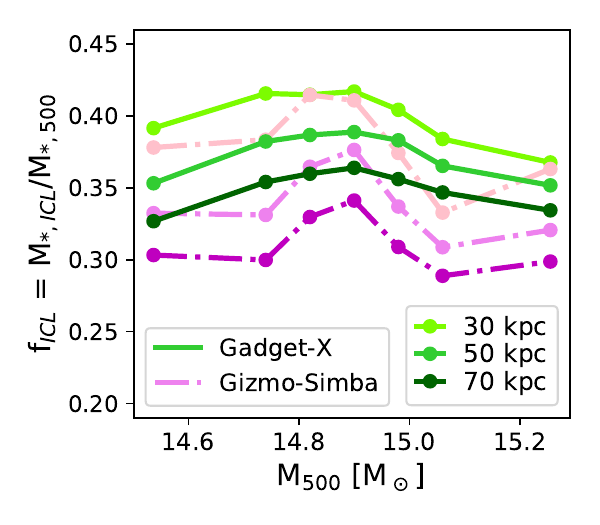}
  \caption{Same as \Fig{fig:massfraction}, ICL mass fraction as a function of cluster mass, but comparing the values for different aperture sizes to define the BCG, in solid green for \textsc{Gadget-X} and in dash-dotted magenta for \textsc{Gizmo-Simba}. Lighter to darker colours indicate increasing size of aperture: 30, 50 and 70 kpc. We only show the values considering overdensity 500.}
\label{fig:massfraction_apertures}
\end{figure}

Here we repeat the calculations presented in Sec.~\ref{sec:icl_frac} for the different fixed apertures considered. In \Fig{fig:massfraction_apertures} we show the dependence of the ICL mass fraction on cluster mass. This is the same as \Fig{fig:massfraction} but including also 30 and 70 kpc and, for simplicity, showing only the median values for each mass bin, without the points themselves and the errors. As before, green depicts \textsc{Gadget-X} results, while magenta is for \textsc{Gizmo-Simba}, while lighter to darker colours represent smaller to larger apertures, namely 30, 50 and 70 kpc. Although the values change as expected, that is, a smaller BCG makes a larger ICL, we can see that the same trends remain, with no correlation between the ICL mass fraction and the cluster mass for our considered mass range. We also show only the results for overdensity 500, but note that the same trends remain also for $R_{200}$.

\begin{figure}
   \centering
   \includegraphics[width=6.7cm]{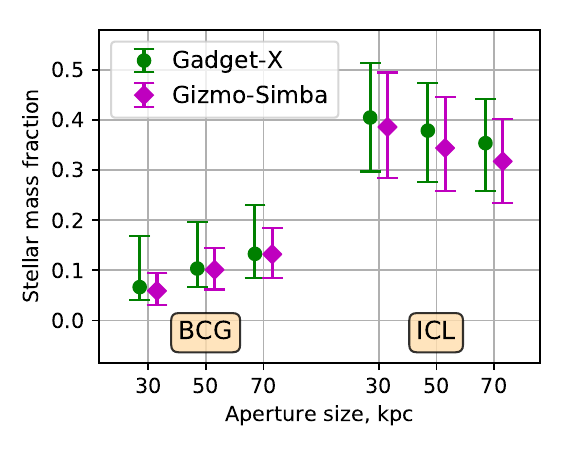}
   \caption{Stellar mass fraction for the BCG and ICL comparing the aperture size used to define the BCG. This is computed dividing the stellar mass content in each of the components by the total stellar mass inside $R_{500}$, $M_{*,500}$. The dots show the median value for the 324 clusters, while the error bars indicate 16th-84th percentiles. Green circles are for \textsc{Gadget-X}, magenta diamonds for \textsc{Gizmo-Simba}.}
\label{fig:fractions_barplot_apertures}
\end{figure}

In \Fig{fig:fractions_barplot_apertures}, we show the stellar mass fraction in the two separate components BCG and ICL. This is the fraction of the total stellar mass that belongs to these components, computed as $f_\mathrm{BCG,500} = M_\mathrm{BCG}/M_{*,500}$ or the same for $f_\mathrm{ICL}$. Unlike in \Fig{fig:fractions_barplot} (see \Sec{sec:icl_frac}), here we do not show the values for the satellite galaxies, that is stars that do not belong to the BCG neither the ICL, because this value is independent of the aperture size used to define the BCG, which does not affect the joint component BCG+ICL. This figure is similar to \Fig{fig:massfraction_apertures}, but in \Fig{fig:fractions_barplot_apertures} we focus on the median values rather than the dependece on cluster halo mass, and we also present results for the BCG mass fraction. We can see that the median value of $f_\mathrm{BCG}$, which was around 10 per cent for 50 kpc, decreases to $\sim 7$ per cent for 30 kpc and increases to $\sim 13$ per cent for the largest aperture of 70 kpc. For the ICL, since the absolute values are larger, the relative changes from one aperture to the other are smaller, with the same trends as for the BCG. We do not show here the fractions when selecting only the most relaxed and disturbed clusters (see \Fig{fig:fractions_barplot_sep}), but we note that the trends remain the same regardless of the aperture used.

\subsection{Relation of ICL to dark matter}

\begin{figure*}
   \centering
   \includegraphics[width=11cm]{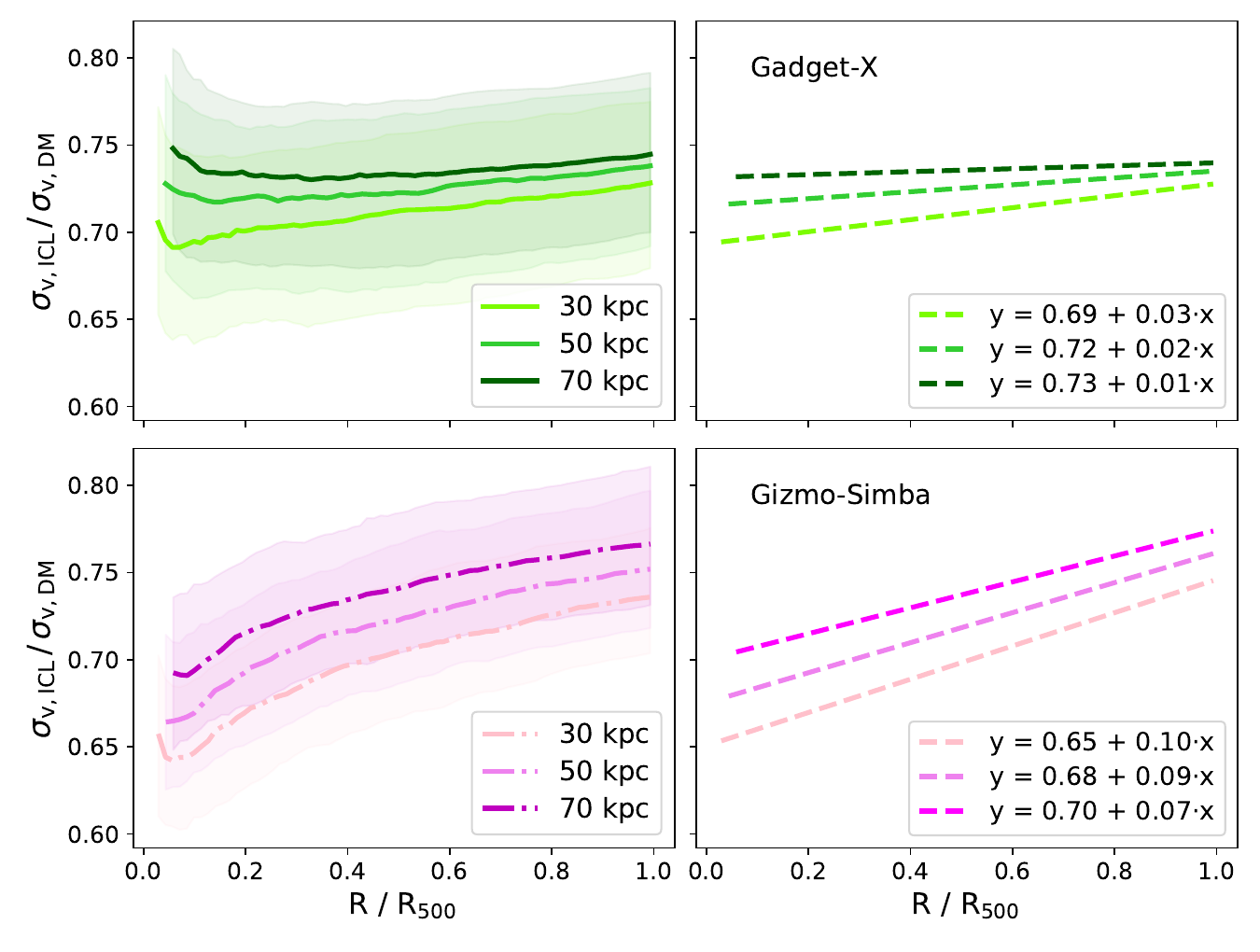}
   \caption{Dependence of the relation between ICL and DM profiles on the aperture size used to define the BCG. First column: ratio between the ICL and DM velocity dispersion profiles, comparing the results for the three different apertures considered to define the BCG and thus the ICL: 30, 50 and 70 kpc. In green for \textsc{Gadget-X} and in magenta for \textsc{Gizmo-Simba}. Second column: fit of the previous plots to a straight line, with the equations of these lines indicated in the legend.}
\label{fig:icldm_fits_apertures}
\end{figure*}

We also repeat now the results of \Sec{sec:icl_dm}, where we showed the relation between ICL and DM in the clusters, by comparing their mass density and velocity dispersion profiles (see \Fig{fig:icldm_fits}). For the density profile, changing the aperture does not affect the profile except by how far in the inner part the profile gets, the shape remains always the same. This is because of the way the density profile is constructed, and thus there is no reason to repeat it for different sizes of the aperture. However, for the velocity dispersion, the aperture size can change the results, and hence we are depicting this change in \Fig{fig:icldm_fits_apertures}. In this plot, the left column shows the median values of the ratio of the two profiles, $\sigma_\mathrm{v,ICL}/\sigma_\mathrm{v,DM}$, with lighter to darker colours being for 30, 50 and 70 kpc respectively. The shaded regions are the 16th-84th percentiles. The upper row, in green, is for \textsc{Gadget-X}, while the lower one, in magenta, is for \textsc{Gizmo-Simba}. We can see that this ratio does not change very significantly from one aperture to another. The main effect is that, for larger apertures, the ratio is increased, which means that the ICL profile is higher (the DM profile is not affected by the aperture size). This can be explained by the fact that particles in the innermost region, selected only for smaller apertures, have velocities that are more similar within each other, such that the velocity dispersion profile decreases towards the cluster centre.  

The right column in \Fig{fig:icldm_fits_apertures} shows the fit of these ratios to a straight line with the form $y = c+d \cdot x$. The values of these parameters can be seen in the legend of the plot, as well as summarised together with their 1 $\sigma$ errors in \Tab{table:fit_params_apertures}. As it could be expected from the left column, the lines are very similar between each other, showing only very slight changes. Moreover, in \Tab{table:fit_params_apertures} we can see that, for 30, 50 and 70 kpc, the values of the $c$ and $d$ parameters are within the 1 $\sigma$ intervals of each other, so that they are all compatible. Hence, we can say that choosing a different size of the fixed aperture that encloses the BCG does not make a significant difference regarding the relation between ICL and DM velocity dispersion profiles. 

\begin{table}
\centering
\caption{Parameters of the best fit lines to the ratio between the ICL and DM velocity dispersion profiles shown in \Fig{fig:icldm_fits_apertures}, together with their 1 $\sigma$ errors.}
\begin{tabular}{r c c c}
\hline
    \multicolumn{2}{c}{$y = c + d \cdot x$} &  {\textbf{\textsc{Gadget-X}}} & {\textbf{\textsc{Gizmo-Simba}}}    \\ \hline
  $\bm{c}$  & \textbf{30 kpc}  & $0.69 \pm 0.01$     &  $0.65 \pm 0.01$    \\ 
            & \textbf{50 kpc}  & $0.72 \pm 0.01$     &  $0.68 \pm 0.01$     \\ 
            & \textbf{70 kpc}  & $0.73 \pm 0.01$     &  $0.70 \pm 0.01$     \\ \hline
  $\bm{d}$  & \textbf{30 kpc}  & $0.03 \pm 0.02$     &  $0.10 \pm 0.02$     \\ 
            & \textbf{50 kpc}  & $0.02 \pm 0.02$     &  $0.09 \pm 0.02$     \\ 
            & \textbf{70 kpc}  & $0.01 \pm 0.02$     &  $0.07 \pm 0.02$     \\ \hline
\end{tabular}
\tablefoot{The best fit lines have equation $y = c + d \cdot x$, where $d$ is the value of the slope. We compare, for the two codes \textsc{Gadget-X} and \textsc{Gizmo-Simba}, the results for different BCG aperture sizes.}
\label{table:fit_params_apertures}
\end{table}


\label{lastpage}
\end{document}